\documentclass[twocolumn]{aastex63}

\newcommand{\Mpch}{$h^{-1}$ Mpc}
\newcommand{\Msun}{$\mathrm{M}_{\odot}$ $h^{-1}$}
\newcommand{\ergs}{erg s$^{-1}$}
\newcommand{\Mbh}{$M_{\rm{BH}}$}

\shorttitle{Local AGN-halo connection}
\shortauthors{Powell et al.}

\graphicspath{{./}{figures/}}
\usepackage{amsmath,amstext}

\begin{document}

\title{BASS. XXXVI. Constraining the Local Supermassive Black Hole - Halo Connection with BASS DR2 AGN}

\author[0000-0003-2284-8603]{M. C. Powell}
\affiliation{Kavli Institute for Particle Astrophysics and Cosmology, Stanford University, 452 Lomita Mall, Stanford, CA 94305, USA}
\affiliation{Department of Physics, Stanford University, 382 Via Pueblo Mall,
Stanford CA 94305}
\email{mcpowell@stanford.edu}

\author[0000-0003-0667-5941]{S. W. Allen}
\affiliation{Department of Physics, Stanford University, 382 Via Pueblo Mall,
Stanford CA 94305}
\affiliation{SLAC National Accelerator Laboratory, 2575 Sand Hill Road, Menlo Park,
CA 94025}
\affiliation{Kavli Institute for Particle Astrophysics and Cosmology, Stanford University, 452 Lomita Mall, Stanford, CA 94305, USA}

\author[0000-0002-9144-2255]{T. Caglar}
\affiliation{Leiden Observatory, PO Box 9513, 2300 RA, Leiden, the Netherlands}

\author[0000-0002-1697-186X]{N. Cappelluti}
\affiliation{Physics Department, University of Miami, Coral Gables, FL 33155}

\author{F. Harrison}
\affiliation{Cahill Center for Astronomy and Astrophysics, California Institute of Technology, Pasadena, CA 91125, USA}

\author[0000-0002-4028-3602]{B. E. Irving}
\affiliation{Kavli Institute for Particle Astrophysics and Cosmology, Stanford University, 452 Lomita Mall, Stanford, CA 94305, USA}
\affiliation{Department of Physics, Stanford University, 382 Via Pueblo Mall,
Stanford CA 94305}

\author[0000-0002-7998-9581]{M. J. Koss}
\affiliation{Eureka Scientific, 2452 Delmer Street Suite 100, Oakland, CA 94602-3017, USA}
\affiliation{Space Science Institute, 4750 Walnut Street, Suite 205, Boulder, CO 80301, USA}

\author[0000-0002-8031-1217]{A. B. Mantz}
\affiliation{Kavli Institute for Particle Astrophysics and Cosmology, Stanford University, 452 Lomita Mall, Stanford, CA 94305, USA}

\author[0000-0002-5037-951X]{K. Oh}
\altaffiliation{JSPS Fellow}
\affil{Korea Astronomy \& Space Science institute, 776, Daedeokdae-ro, Yuseong-gu, Daejeon 34055, Republic of Korea}
\affil{Department of Astronomy, Kyoto University, Kitashirakawa-Oiwake-cho, Sakyo-ku, Kyoto 606-8502, Japan}

\author[0000-0001-5231-2645]{C. Ricci}
\affiliation{N\'ucleo de Astronom\'ia de la Facultad de Ingenier\'ia, Universidad Diego Portales, Av. Ej\'ercito Libertador 441, Santiago 22, Chile}
\affiliation{Kavli Institute for Astronomy and Astrophysics, Peking University, Beijing 100871, People's Republic of China}

\author{R. J. Shaper}
\affiliation{Kavli Institute for Particle Astrophysics and Cosmology, Stanford University, 452 Lomita Mall, Stanford, CA 94305, USA}

\author[0000-0003-2686-9241]{D. Stern}
\affiliation{Jet Propulsion Laboratory, California Institute of Technology, 4800 Oak Grove Drive, MS 169-224, Pasadena, CA 91109, USA}

\author[0000-0002-3683-7297]{B. Trakhtenbrot}
\affiliation{School of Physics and Astronomy, Tel Aviv University, Tel Aviv 69978, Israel}

\author[0000-0002-0745-9792]{C. M. Urry}
\affiliation{Yale Center for Astronomy \& Astrophysics and Department of Physics, Yale University, P.O. Box 208120, New Haven, CT 06520-8120, USA}

\author{J. Wong}
\affiliation{Kavli Institute for Particle Astrophysics and Cosmology, Stanford University, 452 Lomita Mall, Stanford, CA 94305, USA}
\affiliation{Department of Physics, Stanford University, 382 Via Pueblo Mall,
Stanford CA 94305}

\begin{abstract}

We investigate the connection between supermassive black holes (SMBHs) and their host dark matter halos in the local universe using the clustering statistics and luminosity function of active galactic nuclei (AGN) from the Swift/BAT AGN Spectroscopic survey (BASS DR2).  
By forward-modeling AGN activity into snapshot halo catalogs from N-body simulations, we test a scenario in which SMBH mass correlates with dark matter (sub)halo mass for fixed stellar mass. We compare this to a model absent of this correlation, where stellar mass alone determines the SMBH mass.
We find that while both simple models are able to largely reproduce the abundance and overall clustering of AGN, the model in which black hole mass is tightly correlated with halo mass is preferred by the data by $1.8\sigma$. When including an independent measurement on the black hole mass--halo mass correlation, this model is preferred by $4.6\sigma$.
We show that the clustering trends with black hole mass can further break the degeneracies between the two scenarios, and that our preferred model reproduces the measured clustering differences on 1-halo scales between large and small black hole masses.
These results indicate that the halo binding energy is fundamentally connected to the growth of supermassive black holes. 
\end{abstract}


\section{Introduction} 
\label{sec:intro}

It is well established that supermassive black holes (SMBHs) reside at the centers of galaxies and grow during phases of extreme accretion, observed as Active Galactic Nuclei (AGN). While multiwavelength surveys of AGN have made great strides in characterizing the AGN population and their correlated host galaxy properties 
\citep[e.g.,][]{Kormendy:2013,Lanzuisi:2017,Powell:2017,Aird:2018,Suh:2019,Caglar:2020,Koss:2021,Ding:2022}, their connection to their larger-scale environments and host dark matter halos remains largely unconstrained.
There are several proposed mechanisms that drive gas to the galaxy center and trigger AGN activity, which depend on the galaxy's cosmic environments or on the states and/or histories of their host dark matter halos \citep[e.g.,][]{Hopkins:2008,Saha:2013,Galloway:2015,Bower:2017,Marshall:2018,Ricarte:2020}. 
Characterizing the environmental dependence of AGN activity can thus provide powerful constraints on evolutionary scenarios of SMBH fueling and feedback.

AGN clustering measurements have been the primary technique used to determine the host dark matter halo properties of accreting black holes. 
By comparing the well-understood mass-dependent clustering of halos from dark matter simulations to the clustering amplitudes of AGN samples (via the AGN bias parameter), the typical host halo masses (and therefore large-scale environments) of AGN have been estimated, which range from $10^{12} - 10^{13.5}$ \Msun\ depending on the sample \citep[e.g.,][]{White:2012,Eftekharzade:2015,Laurent:2017,He:2018,Timlin:2018,Starikova:2011,Allevato:2011,Allevato:2014,Shen:2009,Krumpe:2012,Coil:2009,Hickox:2009,Cappelluti:2010,Cappelluti:2012,Powell:2020}. However an underlying assumption for these mass estimates is that, in a given survey, AGN are hosted by a narrow distribution of halo masses, since their clustering amplitudes are compared to those of simulated halos in narrow mass bins \citep[e.g.,][]{Tinker:2010}. However, this is not necessarily the case given the wide distributions of accretion rates and the unknown link between SMBH and halo properties \citep{DeGraf:2017,Powell:2020,Aird:2021}. 

An additional complication for clustering measurements is that they can be severely biased by selection effects, since each AGN detection method prefers finding AGN in galaxies of particular masses, star formation rates, and redshifts \cite[e.g.,][]{azadi:2017}. This makes comparisons between various surveys difficult. 
In fact, many recent studies have shown that host-galaxy properties are the primary drivers of AGN clustering \citep{Mendez:2016,Yang:2018,Powell:2018,Powell:2020,Krishnan:2020,Aird:2021} indicating little or no dependence on large-scale environment for AGN activity.
However there have been reports for clustering trends with various AGN parameters, such as obscuration/optical type \citep{Hickox:2011,Jiang:2016,DiPompeo:2017,Powell:2018,Krumpe:2017}, AGN luminosity \citep{Krumpe:2012,Allevato:2014}, and black hole mass \citep{Krumpe:2015}, although some inconsistencies in these results have been shown in different surveys \citep[e.g.,][]{Allevato:2011,Mendez:2016,Powell:2020}.
 Due to the many selection effects at play, forward-modeling the AGN population within cosmological simulations has proven to be an effective way of interpreting AGN clustering with full control of the many selection effects \citep{Powell:2018,Powell:2020,Georgakakis:2019,Comparat:2019,Allevato:2021,Aird:2021}.
However, the limited survey statistics have as yet limited the conclusions that can be drawn regarding the intrinsic connections between SMBH properties and large-scale structure.

State-of-the-art AGN surveys are beginning to have the statistics needed to investigate the dependencies of fundamental SMBH parameters on AGN clustering to probe the underlying SMBH-halo connection.  
The {\it Swift}/BAT Spectroscopic Survey (BASS DR2; \citealt{Koss:2017,Koss:2021a,Koss:2021b}) provides the largest and most complete sample of AGN in the local universe to date.
The ultra-hard X-ray selection of the BASS survey and its well-known selection function make it one of the most unbiased spectroscopic samples, as AGN obscured up to the Compton-thick level (i.e., with column densities $N_{H}>10^{24}$ cm$^{-2}$) are able to be detected \citep[e.g.,][]{Ricci:2015}. The extensive follow-up optical spectroscopy has enabled redshifts, black hole masses, and intrinsic luminosities to be estimated for each AGN \citep{Koss:2021a,Koss:2021b,Oh:2021,Mejia-Restrepo:2022}. 
Additionally, the full sky coverage and depth of the survey overlaps with the Two Micron All Sky Survey (2MASS) redshift galaxy survey (2MRS; $z<0.1$), enabling cross-correlation clustering measurements on 1-halo scales ($<1$ \Mpch) with sufficient statistics and as a function of black hole mass. This provides powerful constraints on how SMBHs occupy their host galaxies and halos in the local universe. 

To investigate the role of the dark matter halo in the formation and growth of SMBHs, we constrain how correlated SMBH mass (\Mbh) is with halo mass. This is done by populating SMBHs and AGN activity in halo catalogs from N-body simulations, forward-modeling the BASS selection, and comparing the mock AGN statistics with the BASS clustering measurements (building on the work of \citealt{Cappelluti:2010} and \citealt{Powell:2018}). Our approach assumes that peak (sub)halo mass is the primary halo property connected to SMBHs. Subhalos are halos that have fallen into the virial radius of a larger parent halo, and they are believed to be associated with satellite galaxies; central galaxies, on the other hand, reside at the centers of the parent halos.
It has been shown that a relationship between galaxy stellar mass and peak (sub)halo mass reproduces galaxy clustering fairly well \citep[e.g.,][]{Conroy:2006}, indicating that stellar mass is the primary galaxy property linked to dark matter halos.
In an analogous way, we test the role of black hole mass by
by constraining its connection to (sub)halo mass, assuming a uniform distribution of normalized accretion rates (Eddington ratios; $\lambda_{\rm Edd}$). 
Unlike previous methods for populating AGN mocks into simulations, this approach relies only on the fundamental AGN properties (\Mbh\ and $\lambda_{\rm Edd}$) and does not require an arbitrary duty cycle parameter. Instead, all SMBHs are assumed to accrete at some level.

For the first time, this work tests two extreme assumptions for the correlation (or not) between \Mbh\ and peak (sub)halo mass by examining whether each can reproduce the clustering and abundance of AGN observed in the local universe. The local scaling relation between black hole mass and stellar mass is constrained under each assumption. We additionally compare the model predictions of the clustering dependence on black hole mass to the BASS measurements to further scrutinize the models.

This paper is organized as follows: Section \ref{sec:dat} describes the AGN and galaxy samples used in this analysis; Section \ref{sec:stats} describes the clustering methodology and luminosity function measurements; Section \ref{sec:mods} describes our two empirical AGN halo models that we test against the data; Section \ref{sec:abc} describes the method used to constrain our model parameters; Section \ref{sec:res} and \ref{sec:dis} presents and discusses the results of our analysis; and in Section \ref{sec:sum} we summarize our results and conclusions.
Throughout this paper we assume a flat $\Lambda$CDM Cosmology ($H_{0}=100~h~\rm km\,s^{-1}~\rm Mpc^{-1}$, $h=0.7$, $\Omega_M = 0.3$, $\Omega_\Lambda = 0.7$).

\section{data}
\label{sec:dat}
\subsection{AGN Sample}
The AGN sample is drawn from the second data release of the {\it Swift}/BAT AGN Spectroscopic survey (BASS DR2; \citealt{Koss:2021a,Koss:2021b}), which comprise 858 sources detected by the hard X-ray $14-195$ KeV band via the BAT detector aboard the Neil Gehrels Swift observatory as part of the 70-month catalog \citep{Baumgartner:2013}. 
Optical spectroscopic follow-up has been obtained such that $99.9\%$ of unbeamed AGN outside the galactic plane have spectroscopic redshifts, and $98\%$ have black hole mass estimates. 

Hard X-rays detect both obscured and unobscured AGN, and so the black hole  masses are derived by a range of techniques depending on the spectral characteristics of each source. The majority of objects with broad ($>1000$ $\rm km\,s^{-1}$) emission lines (Type 1s; H$\alpha$, H$\beta$, Mg $_{\rm{II}}$, and/or C $_{\rm{IV}}$) have masses estimated by the FWHM of those lines with uncertainties $\sim 0.3-0.4$ dex. Details of the mass estimations from broad lines are found in \citealt{Mejia-Restrepo:2022}. Sources without broad lines (Type 2s) rely on the bulge velocity dispersions ($\sigma_{*}$) calculated by the absorption features in the optical spectra (the Ca H+K, Mg $_{\rm{I}}$, and/or the Ca $_{\rm{II}}$ triplet), assuming the $M_{\rm{BH}}-\sigma_{*}$ from \citealt{Kormendy:2013}. These mass estimates are described in detail in  and \citep{Koss:2021a} and Caglar et al., {\it in prep}. Uncertainties of masses estimated by this method are $\sim0.35 - 0.5$ dex.

Soft-X-ray ($0.05-200$ keV) observations have also been taken for each BASS AGN by Swift, Chandra, or XMM-Newton, from which column densities and intrinsic X-ray have been calculated. Bolometric luminosities have also been derived from the intrinsic X-ray luminosities \citep{Ricci:2017B}. Finally, stellar masses have been calculated for a subset of sources via overlapping photometry from 2MASS, WISE, and SDSS (see \citealt{Powell:2018} for details).

The BASS sample has some significant advantages for AGN clustering and population studies due to its completeness, large volume, well-known selection function, and extensive multi-wavelength ancillary data. The full details of the DR2 release is described \citealt{Koss:2021b}. Recently, the intrinsic Eddington Ratio Distribution Functions (ERDFs) for type 1s, type 2s, and the total BASS sample have been measured \citep{Ananna:2022}. We use the ERDF of the full sample to create mocks of our AGN sample in cosmological simulations. 

The luminosity function and intrinsic Eddington Ratio Distribution Function calculation used all unbeamed AGN in the DR2 sample with redshifts $0.01<z<0.3$, black hole masses $6.5 < \log_{10}(M_{\rm{BH}}/M_{\odot})<10.5$, and Eddington ratios $-3<\log_{10}(\lambda_{\rm Edd})<1$, totaling 586 AGN \citep{Ananna:2022}. 
For the clustering measurement we followed \citealt{Powell:2018} and further limited the redshift range to $0.01<z<0.1$ to overlap with the 2MRS galaxies, and selected AGN with $L_{X}>10^{42.5}$\ergs (without the \Mbh\ and $\lambda_{\rm Edd}$ restrictions). In total, 724 AGN were used for the correlation function calculation, which had an average redshift of $z=0.04$. 

\begin{figure}
    \centering
    \includegraphics[width=\linewidth]{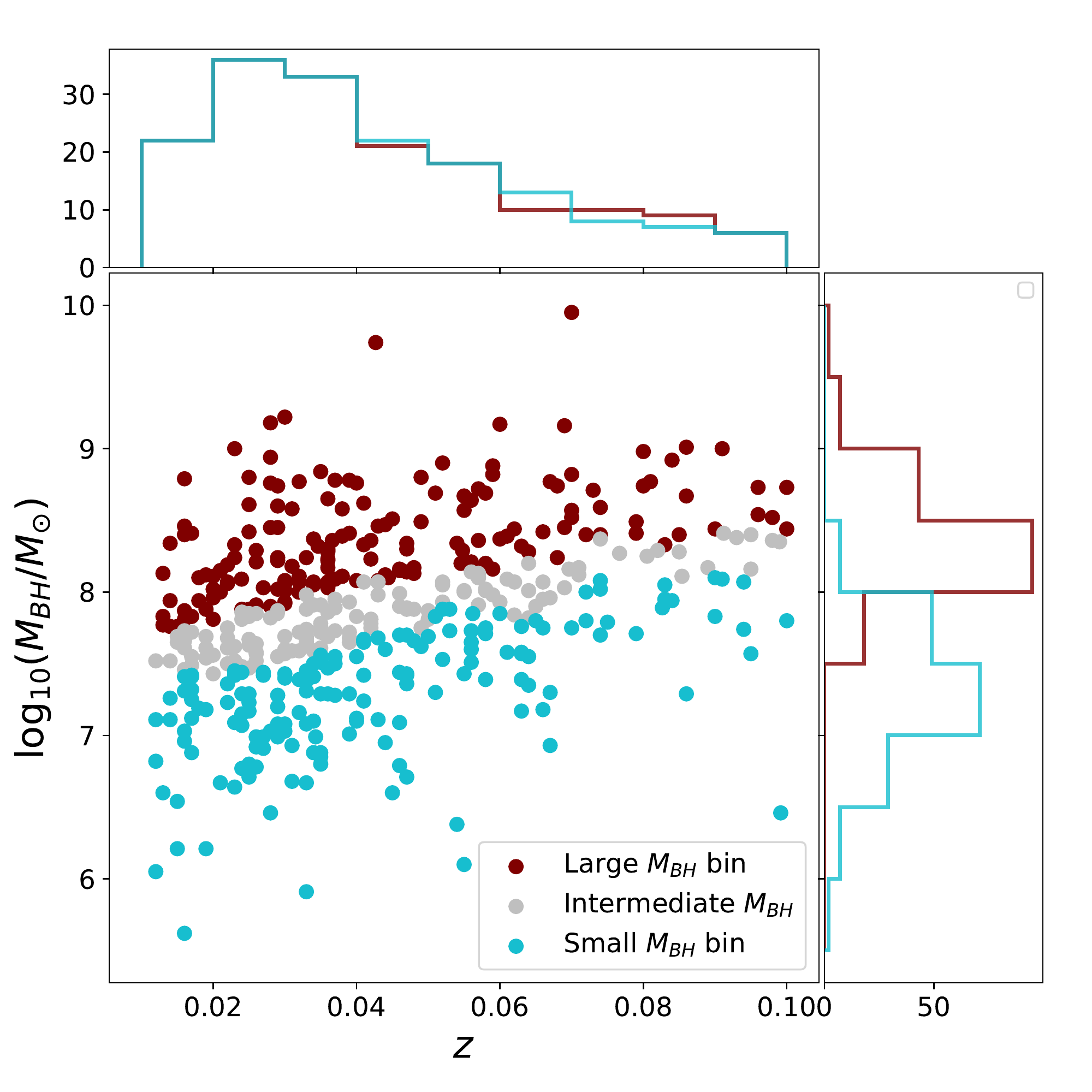}
    \caption{Black hole mass vs. redshift of our AGN sample used in the clustering analysis. We defined two bins of black hole mass (large \Mbh, maroon; small \Mbh, cyan) with similar redshift distributions (top histogram) to investigate clustering trends with black hole mass.}
    \label{fig:mbh_vs_z}
\end{figure}

In addition to the clustering measurement of the full BASS sample, we investigated the clustering dependence on black hole mass (Section \ref{sec:mbh}). For this calculation, we chose two bins of \Mbh\ with the same redshift distributions so that the same volumes were probed for each. This was done by splitting the sample into 10 bins of redshift and selecting the upper and lower third of black hole masses in each bin. We did not use the middle third in order to reduce contamination between the bins, since the uncertainties of our black hole mass estimates are $0.3-0.5$ dex. 
The choice of using the upper and lower $33\%$ of the sample was empirically found as a balance in which the number of AGN used in the measurements was maximized for better statistics, while noise introduced due to bin contamination was minimized. 
We also removed Type 1.9 AGN from each bin, as \Mbh\ estimates have been shown to be biased due to obscuration in these sources \citep{Mejia-Restrepo:2022}. We investigated the impact of using different \Mbh\ measurement methods in Appendix \ref{sec:a1}. 
The \Mbh\ vs $z$ distributions of the full sample, as well as for our two defined \Mbh\ bins, is shown in Figure \ref{fig:mbh_vs_z}.

\subsection{Galaxy Sample}

We selected galaxies from the 2MASS Redshift Survey (2MRS; \citealt{Huchra:2012}) to cross-correlate with our AGN for improved clustering statistics. There are 40308 Ks-band-selected galaxies in the sample with $K_s>11.75$ and $0.01<z<0.1$ that cover 91\% of the sky ($|b|>10^{\circ}$). 
Modeling the galaxy correlation function requires an understanding of the completeness of the sample. This was done by comparing the absolute magnitude distribution of the galaxies with the Ks-band luminosity function \citep{Jones:2006}. We calculated the K-band absolute magnitudes ($M_K$) from the extinction-corrected apparent K-band magnitudes ($K_s$) via the relation
\begin{equation}
    M_K = K_s - 5 \log_{10} \left( \frac{D_L}{10\rm{pc}}\right) - k(z)
\end{equation}
where $D_L$ is the luminosity distance and $k(z)$ is the K-correction. We used k-corrections from \citealt{Bonne:2015}.

\begin{figure*}
    \centering
    \includegraphics[width=.99\textwidth]{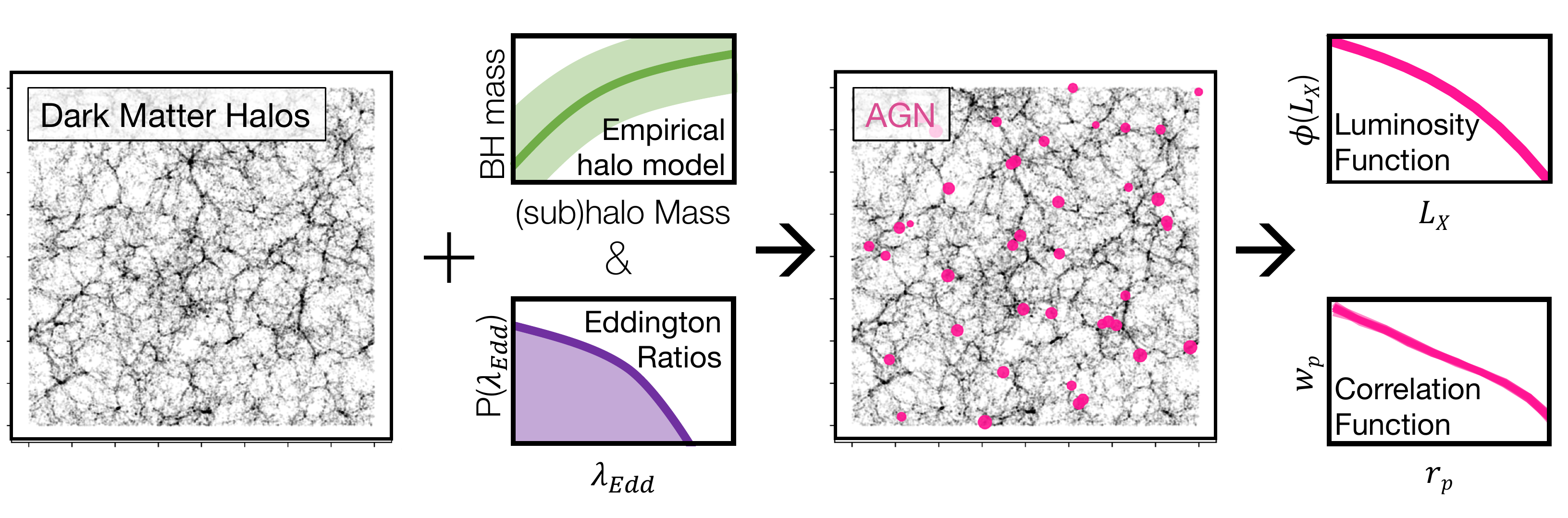}
    \caption{Schematic of our forward-modeling approach to link AGN to their host dark matter halos. We constrain the connection between black hole mass and host (sub)halo mass (“Empirical halo model”) by assuming a universal Eddington ratio distribution function (measured by the data) and comparing the clustering and abundance of mock AGN to the measured X-ray luminosity function and AGN-galaxy cross-correlation of the BASS DR2 sample.}
    \label{fig:diagram}
\end{figure*}

\section{Summary Statistic Measurements}
\label{sec:stats}

Our approach for investigating AGN-halo models is to simulate many mock realizations of the data, compute summary statistics of the mocks and the real data, and then use a quantitative metric of how faithfully the mocks reproduce the real data to constrain the model parameters underlying the mocks.

For the summary statistics, we used the scale-dependent AGN clustering amplitudes and the luminosity-dependent SMBH space densities for the main analysis. 
We describe how each is calculated with the data in the following sections.

\subsection{Clustering Statistics}
The projected two-point AGN-galaxy cross-correlation function is the measure of clustering that we calculate for the data and mock samples. The correlation function is defined as the excess probability that a pair of objects are separated by a given distance scale over a random distribution. In practice, pairs of galaxies are counted in separation bins parallel and perpendicular to the line of sight ($r_p$ and $\pi$, respectively), and compared to the pair counts of a random catalog with the same selection function as the survey. To boost the statistics of the AGN clustering signal, we cross-correlated the AGN with the much more abundant galaxy sample \citep{Powell:2018,Krumpe:2017}. We used the Landy-Szalay estimator \citep{Landy:1993} for this cross-correlation function measurement:

\begin{equation}
    \xi (r_p, \pi) = \frac{AG - AR_G - R_A G - R_A R_G}{R_A R_G},
\end{equation}

\noindent{where each term represents the number of pairs in a separation bin between the AGN ($A$), galaxies ($G$), randoms associated with the AGN survey ($R_A$) and randoms associated with the galaxy survey ($R_G$). Because spectroscopic redshifts determine the $\pi$ measurements, peculiar velocities outside the Hubble flow distort $\xi$ along the $\pi$ dimension. We therefore integrate $\pi$ to average over these redshift-space distortions and obtain the projected correlation function ($w_p$):}

\begin{equation}
    w_p(r_p) = \int_0^{\pi_{max}} \xi(r_p,\pi)d\pi.
\end{equation}

\noindent{We empirically chose $\pi_{max}$ as the value at which $w_p$ plateaus and gets noisier for larger values. This was found to be $40$ \Mpch\ for our sample.}

The AGN random catalogs were generated using the BAT sensitivity map \citep{Baumgartner:2013}. We first randomly assigned angular coordinates within the survey footprint ($|b| >10^{\circ}$) and assigned each random a flux drawn from the log N-log S distribution \citep{Ananna:2022}. We kept only the randoms whose flux values exceeded the survey sensitivity at their respective positions. The angular coordinates for the galaxy randoms were randomly assigned over the 2MRS survey volume excluding the galactic plane ($|b|>10^{\circ}$). The redshifts of each random catalog were drawn from the smoothed redshift distribution of the respective dataset ($\sigma_{z}=0.2$). The AGN random catalog was constructed to be 100 times larger than the BASS DR2 sample; the galaxy random catalog was 20 times larger than the 2MRS sample.

The correlation function uncertainties and covariances were estimated via the jackknife re-sampling technique. The survey was broken up into 25 regions on the sky \citep[e.g.,][]{Powell:2018}, each containing $\sim4 \%$ of the data. We repeated the clustering measurement while excluding data within each region ($w_k$). The number of jackknife samples were chosen so that each patch was large enough to exceed the largest $r_p$ scales at the minimum redshift, but numerous enough to create a normal distribution to estimate the uncertainties.
The covariance matrix was estimated by:

\begin{equation}
\begin{split}
C_{i,j} = \frac{M}{M-1} \sum_{k}^{M} \Big[w_{p,k}(r_{p,i}) - \langle w_{p}(r_{p,i})\rangle\Big]\\
 \times\Big[w_{p,k}(r_{p,j}) - \langle w_{p}(r_{p,j})\rangle \Big] ~~,
\end{split}
\end{equation}

\noindent where $M$ is the number of jackknife samples (25). The errors on $w_p$ for each $r_p$ bin are the square roots of the diagonals:  $\sigma_i=\sqrt{C_{i,i}}$.

\begin{figure*}
\centering
\includegraphics[width=.9\textwidth]{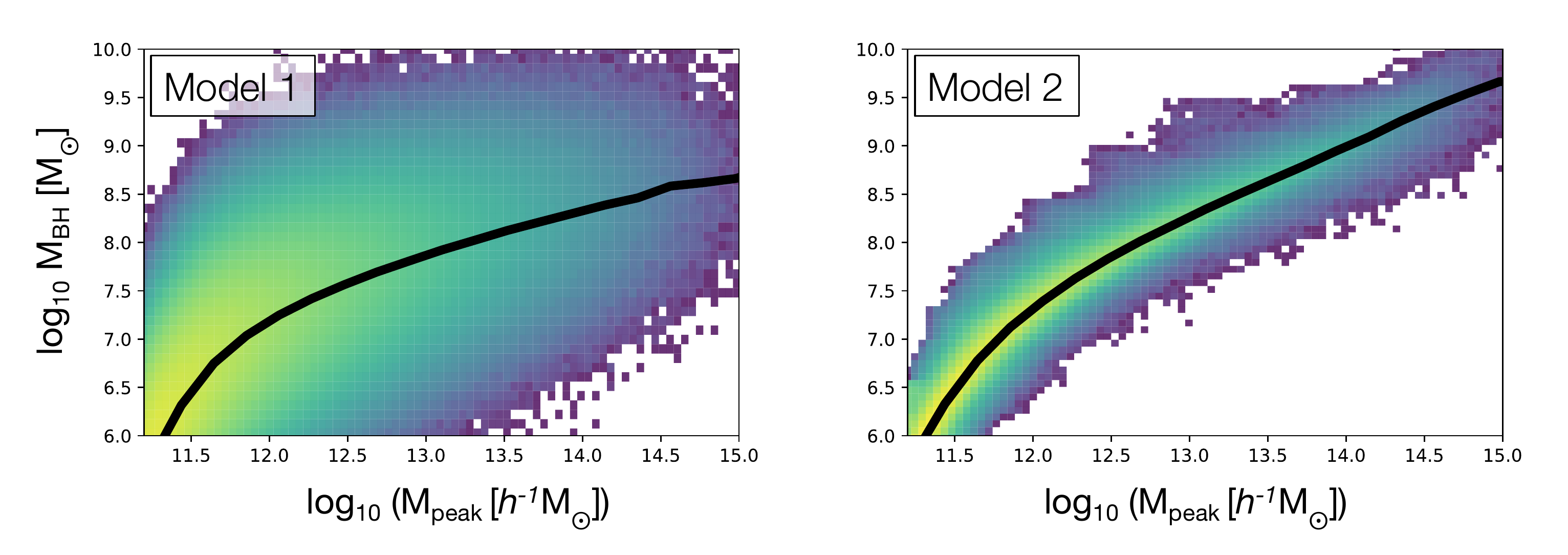}
\caption{2D distributions of the mock supermassive black hole masses (\Mbh) in each model example (Model 1, left; Model 2, right) as a function of host peak (sub)halo mass ($M_{peak}$), shown with a log normal color scale. The median \Mbh$-M_{peak}$ relations are shown by the solid black lines. Model 2 implements a monotonic relation between \Mbh\ and $M_{peak}$ for fixed stellar mass (unlike Model 1), which leads to a strong correlation between SMBH mass and halo mass.} 
\label{fig:mbhdist}
\end{figure*}

\subsection{X-ray Luminosity Function}
The ultra hard (14-195 keV) X-ray luminosity function (XLF) was the secondary summary statistic used as a constraint, which describes the number density of AGN as a function of X-ray luminosity. The BASS XLF was measured via the $1/V_{\rm max}$ method \citep{Ananna:2022}, where each AGN was weighted by the maximum volume that it could have been detected in given its luminosity. 

The uncertainties on the XLF were calculated assuming Poisson statistics, based on the numbers of AGN in each luminosity bin weighted by the $V_{max}$ estimates. See \citealt{Ananna:2022} for details.

\subsection{Independent $M_{\rm BH}-M_{halo}$ Measure}

While the two constraints described above were used for the main analysis, we utilized an independent measurement of the SMBH-halo mass relation from \citealt{Marasco:2021} as a third constraint in Section \ref{sec:third}. \citealt{Marasco:2021} used 55 individual nearby galaxies with dynamically-measured black hole masses, with halo masses inferred either from globular cluster dynamics or spatially resolved rotation curves. They fit a scaling relation to the $M_{\rm BH}-M_{halo}$ correlation: 
\begin{equation}
    \log_{10}\frac{\mathcal{M}_{\rm BH}}{M_{\odot}} = 1.62\times \log_{10}\frac{M_{\rm halo}}{M_{\odot}} - 12.38 ,
\end{equation}
with a scatter of $\sim 0.4$ dex.

While the sample used for this measurement does not fully overlap with our AGN catalog, the correlation provides an independent inference on the SMBH-halo relationship in the local universe.

\section{Empirical subhalo-based models} 
\label{sec:mods}

We used a forward-modeling approach to constrain the AGN-halo connection and interpret the clustering statistics and space densities of the BASS AGN. To summarize, we populated supermassive black holes and galaxies in halo catalogs from snapshots of N-body simulations. We assumed each parent halo and subhalo contains one galaxy and one central black hole, and we assigned masses to each according to empirical models (see below). Each black hole was then assigned an Eddington ratio ($\lambda_{\rm Edd}=L_{bol}/L_{\rm edd}$) drawn from a distribution, which has been directly inferred from the data \citep{Ananna:2022}. A model AGN `mock' sample was then selected with the same AGN luminosities as our BASS sample. In this way, satellite and central AGN were treated the same, with equivalent probabilities for accretion. The abundance and clustering of the mock AGN were compared to the measurements to evaluate the validity of the halo models and their parameters. Figure \ref{fig:diagram} shows a schematic of this approach. 

We used two $z\sim 0$ halo catalogs from two N-body simulations for this analysis. The Small MultiDark Planck simulation (SMDPL; \citealt{SMDPL}) was used when finding the best-fit model parameters via the ABC rejection sampling algorithm (Section \ref{sec:abc}). Its particle mass is $9.6\times 10^7~h^{-1}\rm{M}_{\odot}$, with Planck Cosmology \citep{Planck:2015}. The Rockstar halo-finder \citep{rockstar} was used to obtain the halo catalogs at each redshift. With a box length of $400$ \Mpch, the simulation volume is similar to that of the survey volume. We use the snapshot catalog at scale factor $a=0.971$ (where $a=1/(1+z)$) to match the average redshift of our galaxy and AGN catalog ($z=0.04$). This catalog is used in our ABC rejection sampling algorithm (Section \ref{sec:abc}), as it is large enough for a comparable statistics to our survey but small enough for efficient computation times.

We additionally used the Unit N-body simulation \citep{Chuang:2019}, which has a much larger simulation volume, for an additional consistency check and to plot the best-fit models with improved statistics. With a particle mass of $1.2\times 10^9~h^{-1}\rm{M}_{\odot}$, this simulation also assumed Planck Cosmology and used the Rockstar halo-finder to obtain the halo catalogs. Its box length of $1~h^{-1}$Gpc makes the simulation volume $\sim 10$ times larger than the survey volume. We used the $a=0.978$ snapshot catalog.

\subsection{AGN Halo Models}

For each model, we populated only the halos with virial masses $>5\times 10^{10}~h^{-1}\rm{M}_{\odot}$. Using the \texttt{halotools} software \citep{Hearin:2017},
we assumed the stellar mass-(sub)halo mass relation (SHMR) from \citet{Behroozi:2010} (which includes a log-normal scatter of 0.2 dex at fixed halo mass) to populate each (sub)halo with a mock galaxy of a given stellar mass ($M_*$)\footnote{We explore how our results depend on the SHMR parameters in Appendix \ref{sec:a3}}. Each galaxy was then assigned a black hole mass (\Mbh) according to the host galaxy stellar mass (and sometimes halo mass; see Model 2). We therefore parameterized the models based on the scaling relation between SMBH mass and host galaxy stellar mass, by the normalization and slope of the median relation:

\begin{equation}
\log_{10}(M_{\rm{BH}}/M_{\odot}) = \rm{norm} +  \rm{slope} \times \log_{10}(M_{*}/10^{11}M_{\odot}).
\end{equation}

\noindent{
Log-normal scatter on this relation was included and was an additional parameter constrained; we therefore had three free parameters for each model (normalization, slope, and scatter). There have been previous direct measurements of \Mbh$-M_{*}$ \citep{Kormendy:2013,Reines:2015,Shankar:2016,Suh:2020}; while most measurements of local AGN are consistent, there are mild differences between each reported scaling relation. There is also an offset between the scaling relation of local inactive black holes in ellipticals and the rest of the AGN population, either due to a selection effect \citep{Shankar:2016} or other evolutionary scenarios \citep{Aird:2022}. We therefore left the \Mbh$-M_{*}$ parameters free to determine the best-fit relation in this context.}

The mock black holes were each assigned an Eddington ratio drawn from the Eddington Ratio Distribution Function (ERDF), constrained directly by the data \citep{Ananna:2022} while taking sample incompleteness and truncation into account. We marginalized over the uncertainties in the ERDF parameters by drawing from the parameter priors reported in \citealt{Ananna:2022} for each model realization.

\begin{figure}
\centering
\includegraphics[width=.5\textwidth]{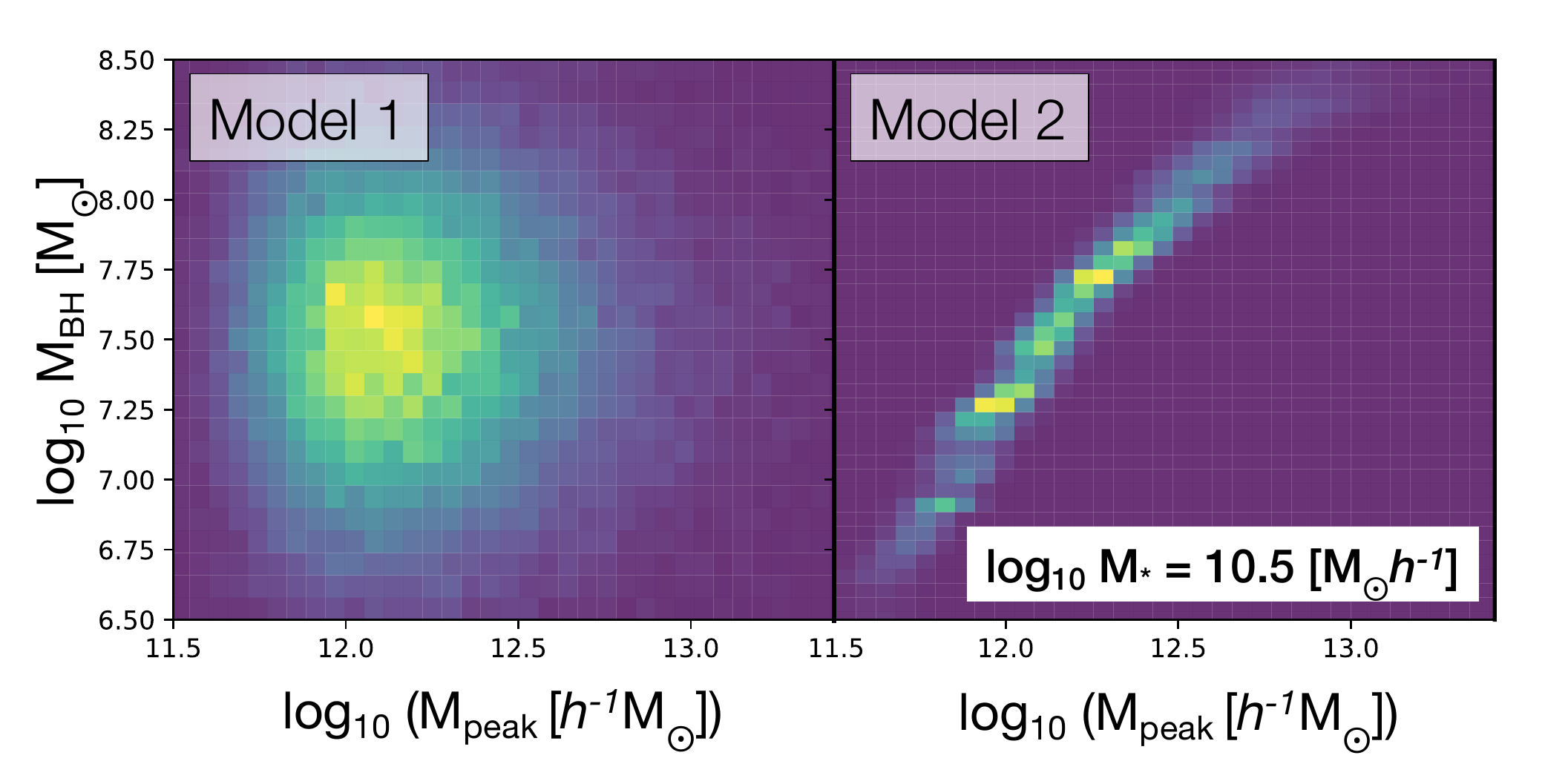}
\caption{2D distributions of the mock supermassive black hole masses (\Mbh) in each model example (Model 1, left; Model 2, right) {\it at fixed stellar mass} $\log_{10}(M_{*}) = 10.5$ [M$_{\odot}$ $h^{-1}$] as a function of host peak (sub)halo mass ($M_{peak}$), demonstrating the different assumptions between \Mbh\ and $M_{peak}$ for each model.} 
\label{fig:mbhdist2}
\end{figure}

The assigned black hole masses and Eddington ratios determined the bolometric AGN luminosities via the relation:
\begin{equation}
    L_{bol} =  1.26\times 10^{38} (M_{\rm{BH}}/M_{\odot}) \lambda_{\rm Edd}~\rm{s^{-1}}\rm{erg}~~
\end{equation}

\noindent{The bolometric luminosities were transformed into hard ($14-195$ keV) X-ray luminosities ($L_{\rm X}$) by assuming a bolometric correction of 8 ($L_{X} = L_{bol}/8$; this corresponds to a 2--10 keV bolometric correction of 20, as found for local sub-Eddington AGN; \citealt{Vasudevan:2007}). 
The luminosity distribution of the mocks was then compared to the data to obtain the incompleteness fraction of the BASS sample as a function of $L_X$. This incompleteness $f$ is defined as  $f(L_X)=N_{dat}/N_{sim} \times V_{sim}/V_{dat}$, where $N_{sim}$ and $N_{dat}$ are the number of mock AGN and BASS AGN in each luminosity bin, respectively, and $V_{sim}$ and $V_{dat}$ are the volumes of the simulation and survey. This was calculated in 10 bins of $\log L_X$ from $43 - 47$ dex [\ergs] and interpolated. We assigned each mock black hole a random value between 0 and 1 and masked out those whose values fell above $f$ for their associated X-ray luminosity. This resulted in a subsample of mock AGN with the same (intrinsic) X-ray luminosity distribution as the BASS AGN.
}

We tested two different assumptions for the additional correlation (or not) between halo mass and SMBH mass. Each is described below. Examples of the resulting distributions of black hole masses as
a function of peak (sub)halo mass are shown in Figure \ref{fig:mbhdist}, showing the different amounts of scatter between black hole mass and halo mass for each.

\subsubsection{Model 1: Simplest Case}
We first assumed a model in which the black hole mass is related to the halo via the SMBH$-$galaxy and galaxy$-$halo relations as previously described, with no additional dependence on large-scale environment or halo properties. We assigned stellar masses to the simulated galaxies via the stellar mass- peak (sub)halo mass relation from \citealt{Behroozi:2010} with 0.2 dex log-normal scatter. Each galaxy hosted a mock supermassive black hole with a mass based on our parameterized $M_{BH}-M_{*}$ scaling relation. 

The clustering and space densities of the resulting mock AGN were compared to the BASS DR2 measurements to constrain the $M_{BH}-M_{*}$ parameters and to test whether this simple model could reproduce the measured survey statistics.

\subsubsection{Model 2: Black hole mass - (sub)halo mass correlation for fixed stellar mass}

We secondly assumed a model in which the black hole mass correlated with the peak mass of its host (sub)halo ($M_{peak}$) for fixed stellar mass. Peak mass was chosen since 'peak' halo parameters have proven to be more successful at interpreting galaxy clustering measurements, as the maximum mass over the history of the halo is not sensitive to mechanisms like stripping that remove dark matter mass at different timescales than baryon mass \citep[e.g.,][]{Behroozi:2014,vandenBosch:2016}. Peak (sub)halo mass ($M_{peak}$) is a standard parameter that subhalo abundance matching techniques (like the one we assumed for this study; \citealt{Behroozi:2010}) link to galaxy stellar mass.

We repeated the above method for populating the mock galaxies within the halos, but additionally introduced this secondary correlation when populating the SMBHs within the galaxies. This was done via the conditional abundance matching technique \citep{Hearin:2013}, which assumes that, for a given stellar mass, the most massive black holes reside in the most massive halos and the least massive black holes reside in the smallest subhalos. This introduces a relation between SMBH mass and peak (sub)halo mass while maintaining the established SMBH$-$galaxy and galaxy$-$halo connections. 

We implemented this technique using the \texttt{conditional abunmatch} function in the \texttt{halotools} software package \citep{Hearin:2017}. The effect of this is demonstrated in Fig. \ref{fig:mbhdist2}, showing the differences between the assumed relations of \Mbh\ and $M_{peak}$ at fixed stellar mass for each model. As before, the clustering and space densities of the resulting mock AGN were compared to the BASS DR2 measurements, and the $M_{BH}-M_{*}$ parameters were constrained.

\subsection{Galaxy Halo Model}
A mock galaxy sample was required for the model cross-correlation function measurement. For this purpose, we simulated the 2MRS galaxies to cross-correlate with our AGN mocks.

We generated the galaxy mocks within the same halo catalog as the mock AGN. A relation between $K-$band luminosity and 
(sub)halo peak mass was assumed, which we modeled via subhalo abundance matching. Using the luminosity function from \citealt{Jones:2006} that was derived from 2MASS galaxies, we used the \texttt{abundancematching} python code\footnote{https://github.com/yymao/abundancematching} to match the abundances of the 2MRS galaxies and simulated dark matter halos assuming a scatter of 0.5 dex (the scatter is larger than in the AGN models since luminosities are used rather than masses). 

After populating the $M_{vir}>5\times 10^{10}~$\Msun\ halos with galaxies and assigning each a K-band luminosity via the resulting relation, we then compared the mock galaxy K-band magnitude distribution of that of the data to obtain the incompleteness fraction as a function of $M_K$. Similar to the AGN, we calculated the incompleteness via $f(M_K)=N_{dat}/N_{sim} \times V_{sim}/V_{dat}$, where $N_{sim}$ and $N_{dat}$ are the numbers of mock galaxies and 2MRS galaxies in each magnitude bin, respectively, and $V_{sim}$ and $V_{dat}$ are the volumes of the simulation and survey. This was calculated in 50 bins of $M_K$ ranging from $-27$ to $-19$. We assigned each mock galaxy a random value between 0 and 1 and masked out those whose values exceeded $f$ for their magnitude. This resulted in a sample of mock galaxies with the same K-band luminosity distribution as the 2MRS galaxies. A mock galaxy sample was generated for each halo catalog used (SMDPL and Unitsim), and we verified that their autocorrelation functions were consistent with the measured 2MRS autocorrelation function.

\begin{figure*}
\centering
\includegraphics[width=.49\textwidth]{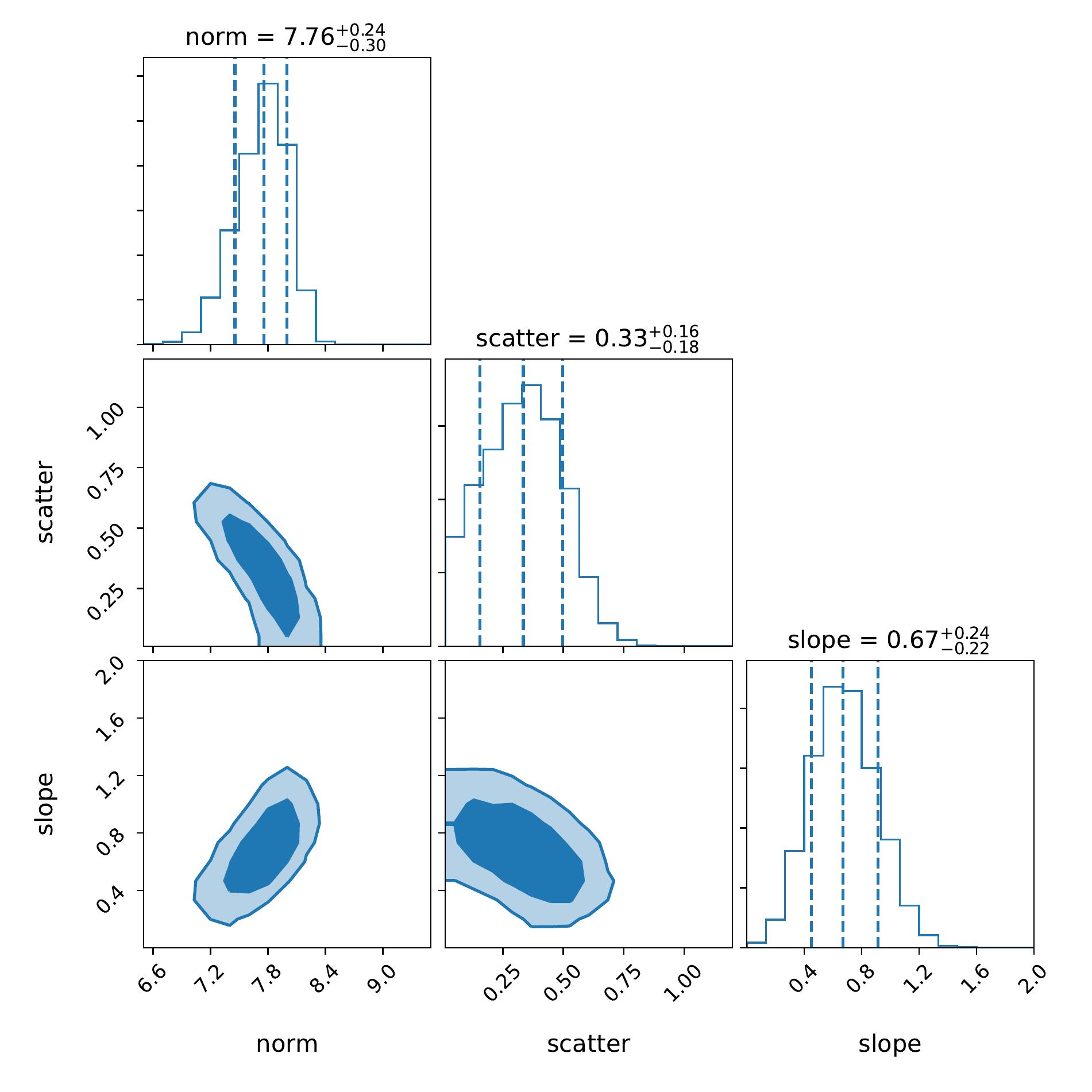}
\includegraphics[width=.49\textwidth]{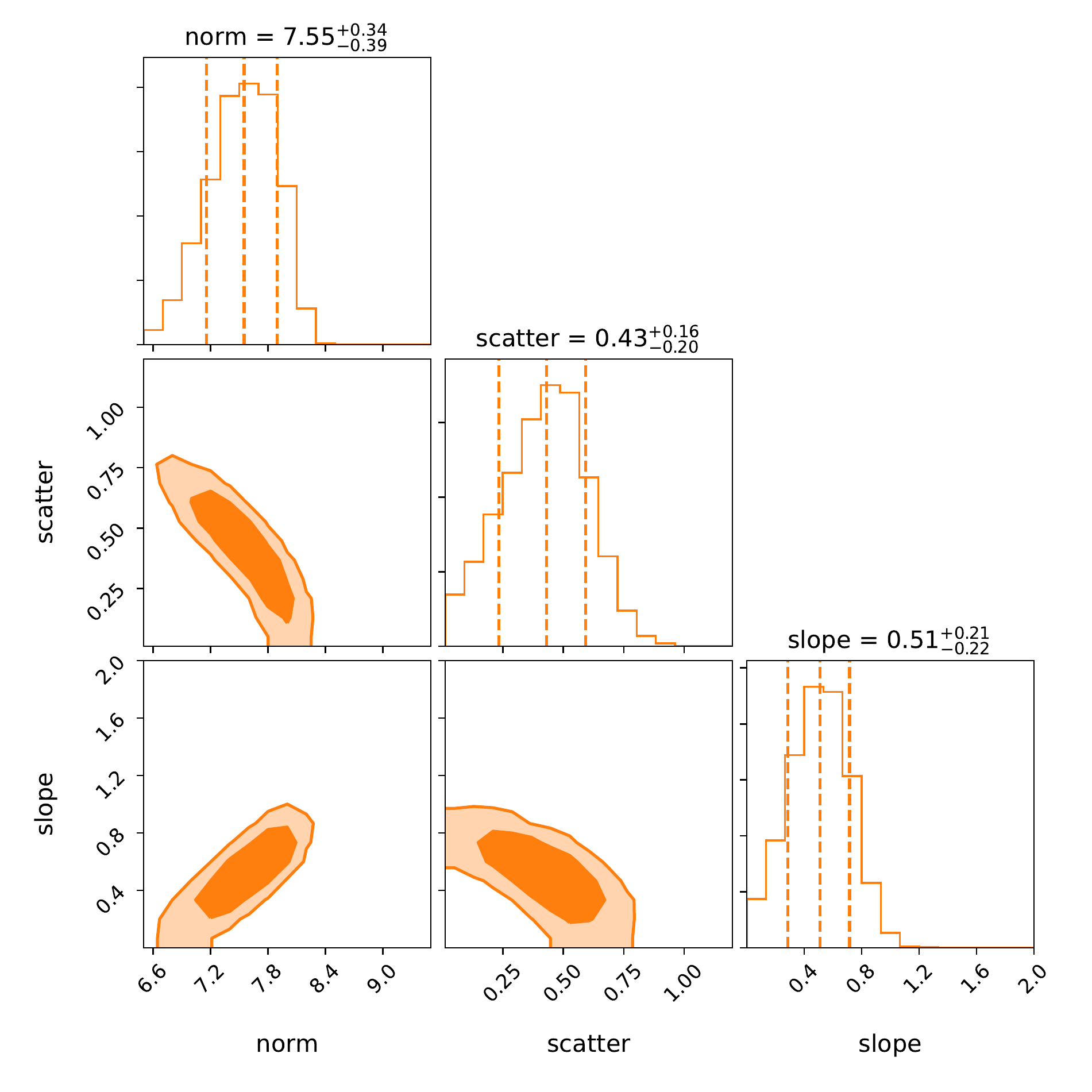}
\caption{Constraints on the $M_{BH}-M_{*}$ relation for Model 1 (left; blue) and Model 2 (right; orange) based on the BASS DR2 AGN-galaxy cross-correlation function and X-ray luminosity function. The diagonal panels plot the posterior distributions of each parameter, where the dashed vertical lines correspond to the $16^{th}$, $50^{th}$ and $84^{th}$ percentiles. The constraints on the normalization, scatter, and slope of the SMBH-galaxy scaling relation are consistent between both models.} 
\label{fig:posts}
\end{figure*}

\section{Approximate Bayesian Computation}
\label{sec:abc}

To fit the parameters of the empirical AGN halo models we employed the Approximate Bayesian Computation (ABC) approach, which has been applied in astrophysical contexts in recent work \citep[e.g.,][]{Hahn:2017,Simola:2019}. This method utilizes rejection sampling to estimate the joint probability for a simulated mock AGN sample given the parameters of the model and a set of observed data.

The main advantage of the ABC method over other traditional MCMC approaches is that the assumption of a Gaussian likelihood function is relaxed. Alternatively, ABC takes advantage of the fact that the probability of observing the data given a simulated model is proportional to the probability of the 'distance' between the data and the simulation being less than an arbitrarily small threshold.

Specifically, we drew parameter proposals given a prior, simulated the AGN data, computed the summary statistics, and then compared them with the observational measurements. If the distance $d$ between the summary statistics of the simulation and data was above the defined threshold, then the parameter proposal was rejected. Otherwise, it was kept and the resulting accepted parameter distributions approximated the posteriors.

In this case there were two primary summary statistics that we compared between the model and data, namely, the cross-correlation function and luminosity function. Following \citealt{Hahn:2017}, we used a multivariate distance measure for our analysis $d = [\rho_{w},\rho_{\phi}]$, where

\begin{equation}
    \rho_{w} = \sum_i  \frac{(w^{m}_{p}(r_{p,i}) - w^{d}_{p}(r_{p,i}))^2}{\sigma_{w,i}^2} 
\end{equation}
\begin{equation}
    \rho_{\phi} = \sum_{i} \frac{(\phi^{m}_i - \phi^{d}_i)^2}{\sigma_{\phi_{i}}^2} .
\end{equation}

\noindent
The quantities $w^{m}_{p}$ and $w^{d}_{p}$ correspond to the model and data projected correlation function, respectively, and $\phi^{m}$ and $\phi^{d}$ are the model and data luminosity functions. $\sigma_{w}^{2}$ and $\sigma_{\phi}^2$ correspond to the diagonal elements of the covariance matrix of the correlation function and luminosity function, respectively.

In Section \ref{sec:third}, we utilize a third constraint: an independent measurement of the SMBH-halo mass relation ($\mathcal{M_{\rm BH}}(M_{\rm halo})$) from \citealt{Marasco:2021}. For this analysis we used the distance measure $d = [\rho_{w},\rho_{\phi},\rho_{\mathcal{M}}]$, where
\begin{equation}
    \rho_{\mathcal{M}} = \sum_{i} \frac{(\mathcal{M}^{m}_i - \mathcal{M}^{d}_i)^2}{\sigma_{\mathcal{M}_{i}}^2} 
\end{equation}

and $\sigma_{\mathcal{M}}$ is 0.4 dex.

Using a Population Monte Carlo algorithm \citep{Hahn:2017,Simola:2019}, we iteratively decreased the threshold until the parameter constraints stabilized. We started by accepting all parameter proposals drawn from the prior. We then calculated the average distance measures of the proposals and defined this average as the 2D threshold for the next iteration. New proposals were made, and the subset whose distances were below the threshold was accepted. The average of the accepted distance measures became the new threshold for the next iteration, and new parameter proposals were drawn from the smoothed distribution of accepted values. This was repeated until the uncertainties on the best-fit model parameters converged and showed negligible changes for any additional iterations. We found that 7 iterations was sufficient for convergence, and we repeated the entire process 100 times to ensure that there were no biases due to the probabilistic nature of drawing from the previous distribution of accepted parameters. We used 10000 parameter draws for each iteration.

The following uniform priors were assumed for our 3 free parameters of the $M_{\rm BH}-M_{*}$ relation: $6.5-9.5$ dex [$M_{\odot}$] for the normalization, $0-1.2$ dex for the scatter, and $0-2$ for the slope.

\begin{figure}
    \centering
    \includegraphics[width=.49\textwidth]{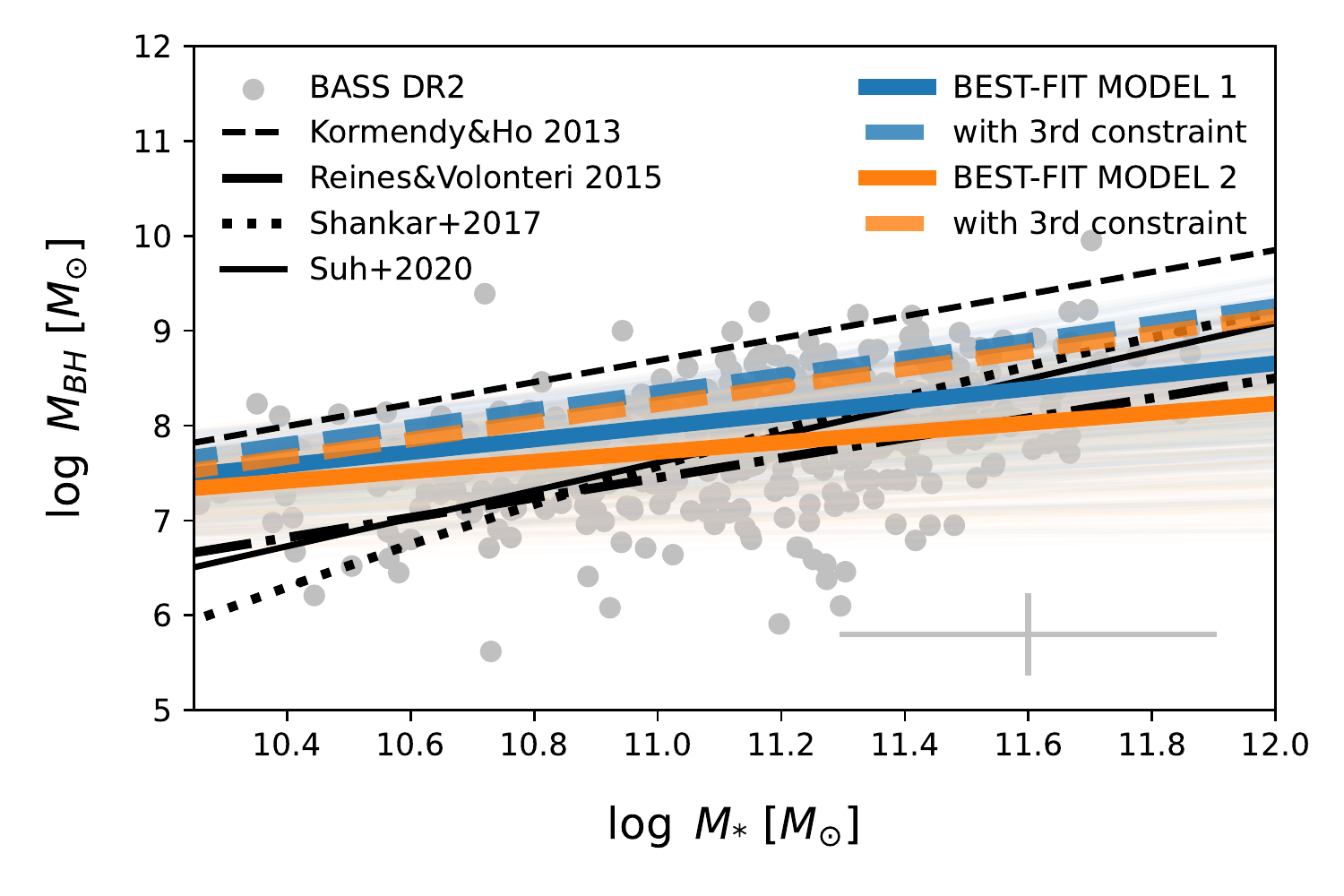}
    
    \caption{Best-fit $M_{\rm{BH}}-M_{*}$ relations from ABC results (blue and orange lines for Model 1 and 2, respectively) using only the full BASS correlation function and luminosity function as constraints. The dashed lines correspond to the relations predicted by the best fit models when imposing the 3rd constraint (Section \ref{sec:third}). Our results are compared to the subset of BASS data with stellar mass estimates (gray circles), as well as previously reported relations from direct measurements (black lines;
    \citealt{Kormendy:2013,Reines:2015,Shankar:2016, Suh:2020}). Typical errors on the mass measurements from the data are shown by the gray cross in the bottom right-hand corner.}
    \label{fig:mbh-ms}
\end{figure}

\begin{figure*}
\centering
\includegraphics[width=.49\textwidth]{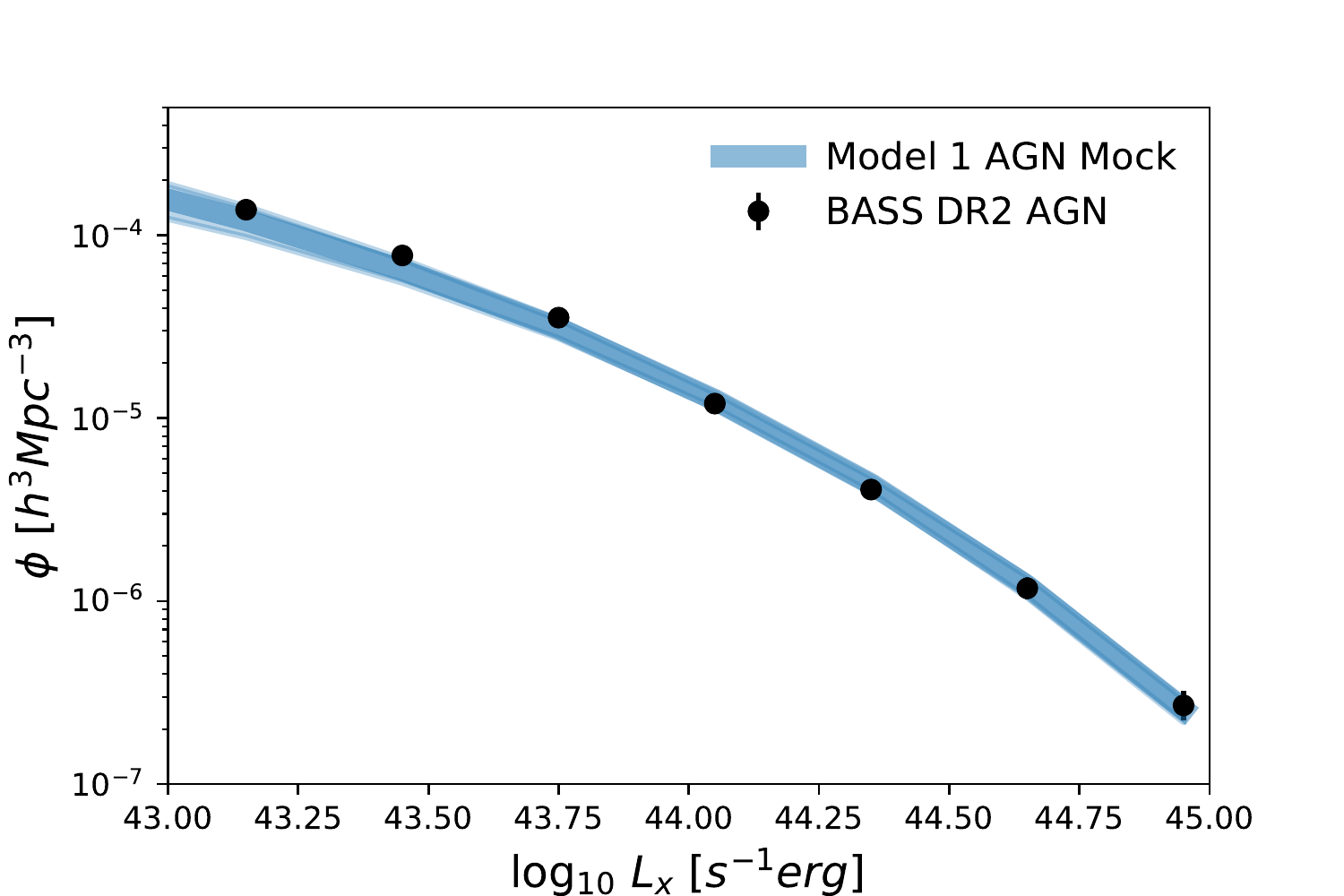}
\includegraphics[width=.49\textwidth]{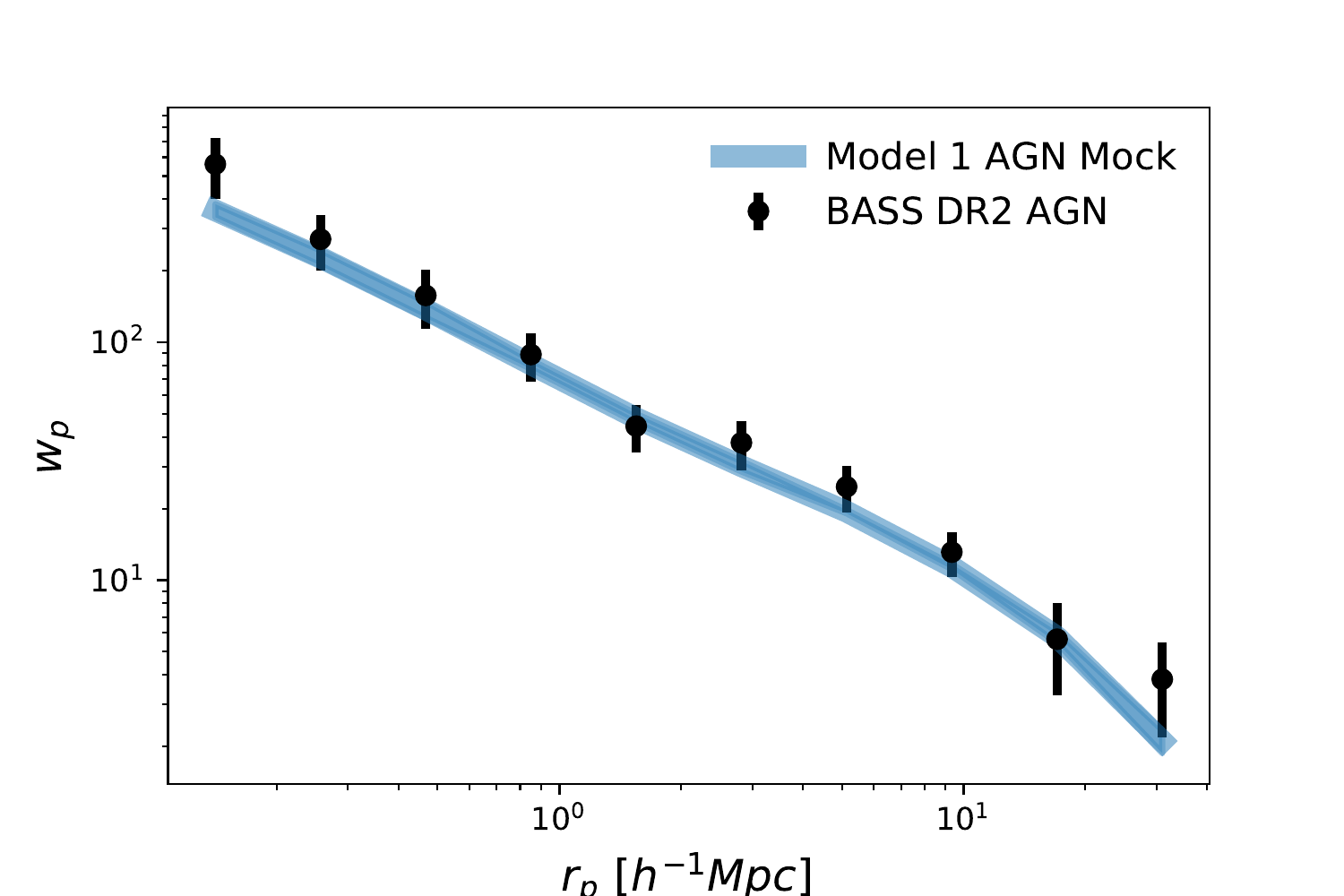}
\caption{Summary statistics of BASS survey (black data points) compared to the best-fit Model 1 mock AGN (blue lines). To within error, the X-ray luminosity function (left) and cross-correlation function (right) are consistent between the model and data despite the very simple model assumptions.} 
\label{fig:stats1}
\end{figure*}

\begin{figure*}
\centering
\includegraphics[width=.49\textwidth]{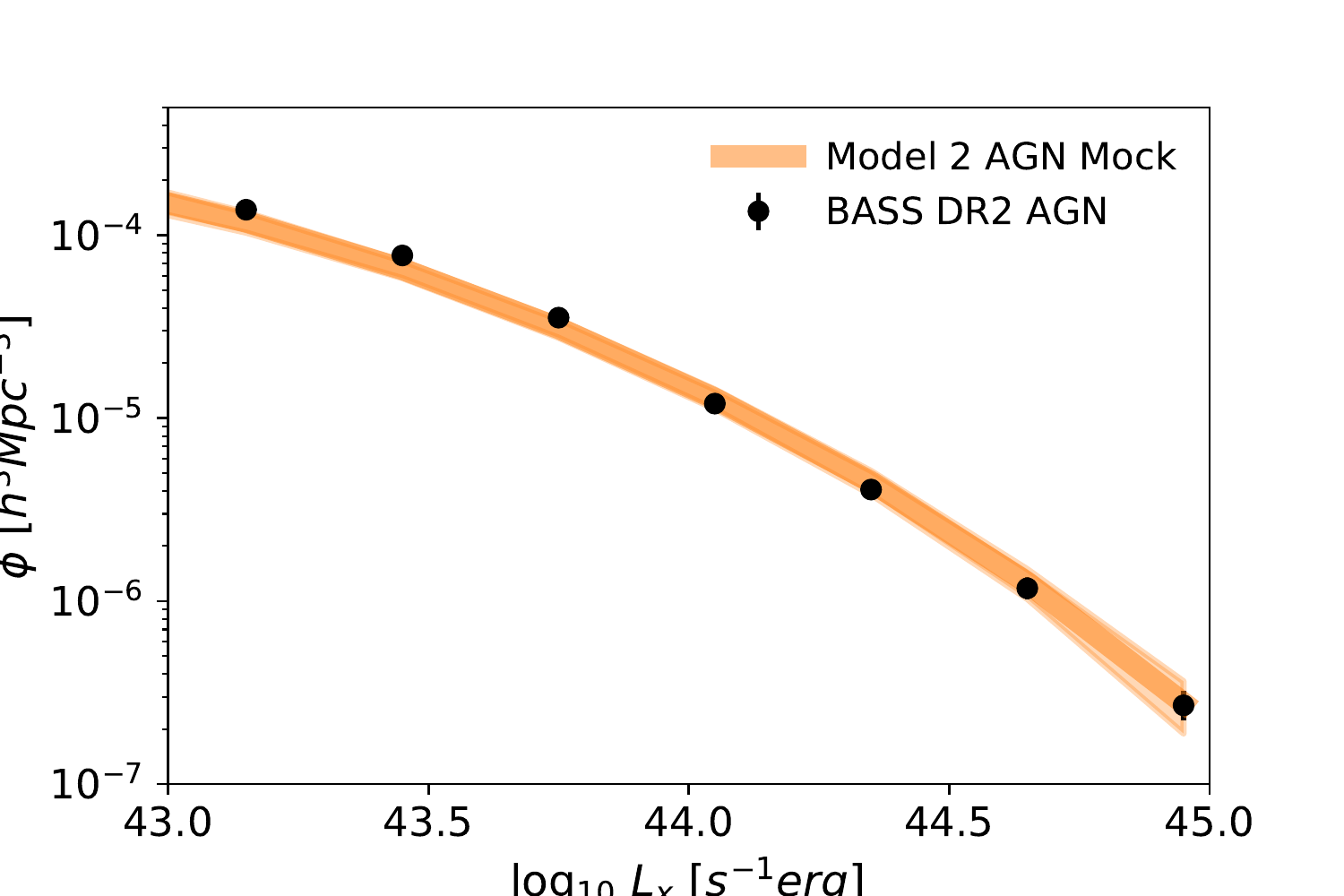}
\includegraphics[width=.49\textwidth]{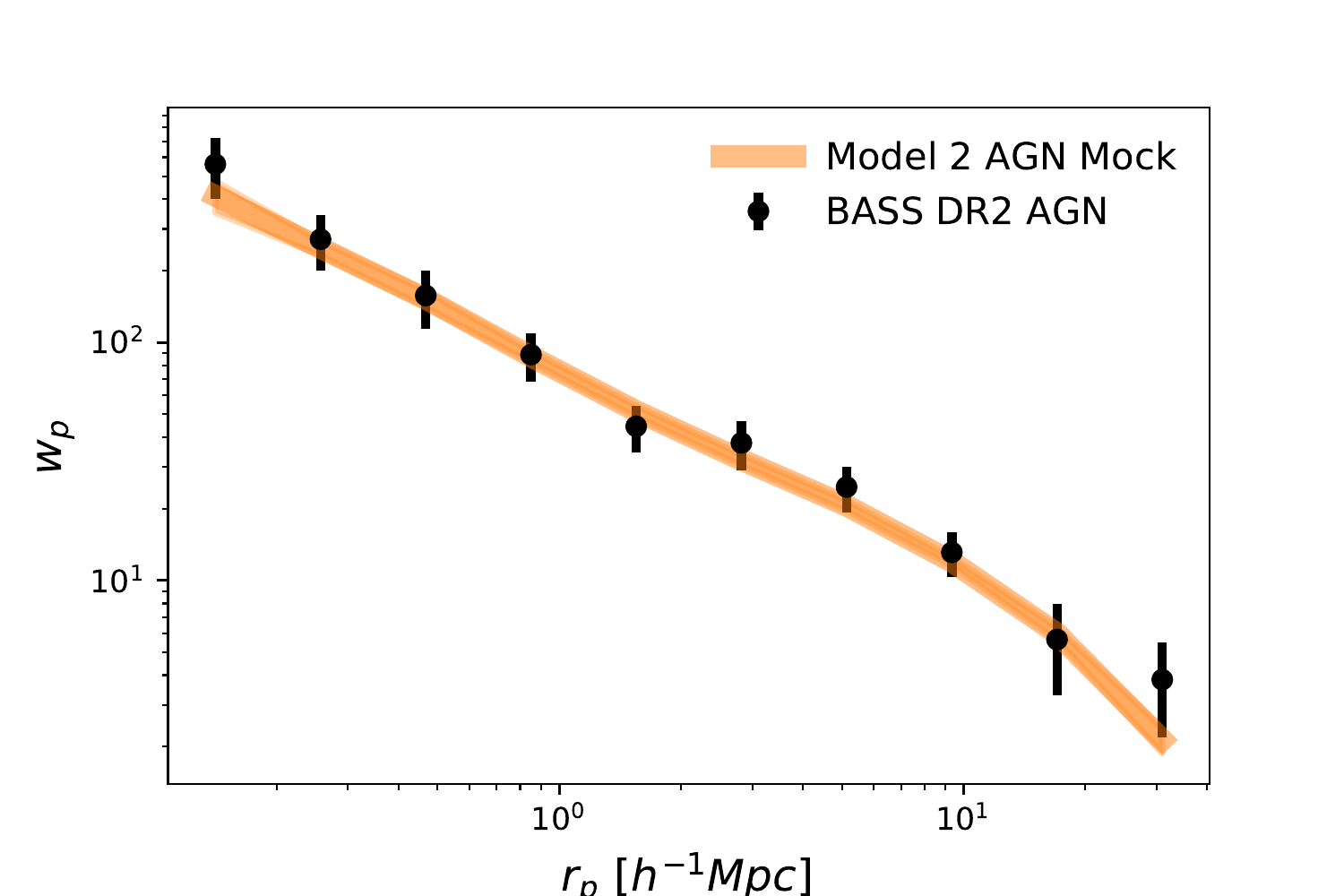}
\caption{Summary statistics of BASS survey (black data points) compared to the best-fit Model 2 mock AGN (orange lines). The model X-ray luminosity function (left) and cross-correlation function (right) are well-matched to the measurements. The Bayes Factor ($P(M_2|D)/P(M_1|D)=1.30\pm0.18$) indicates that this model is preferred by the data by $1.75\sigma$.} 
\label{fig:stats2}
\end{figure*}

\section{Results}
\label{sec:res}

Figure \ref{fig:posts} shows the results from our ABC method to constrain the $M_{\rm{BH}}-M_{*}$ parameters. The left-hand panel shows the posteriors for Model 1; we found that the best-fit \Mbh$-M_{*}$ relation has a normalization of $7.76^{+0.24}_{-0.30}$ dex [\Msun], an intrinsic scatter of $0.33^{+0.16}_{-0.18}$ dex, and slope of $0.67^{+0.24}_{-0.22}$. The right-hand panel of Figure \ref{fig:posts} shows the results for Model 2, which imposes a secondary correlation between \Mbh\ and $M_{peak}$. We found that the best-fit parameters of the \Mbh$-M_{*}$ relation have a normalization of $7.55^{+0.34}_{-0.39}$, an intrinsic scatter of $0.43^{+0.16}_{-0.20}$, and a slope of $0.51^{+0.21}_{-0.22}$. The constraints on these parameters are consistent with Model 1 (Fig. \ref{fig:mbh-ms}). 
Note that these parameters depend on the assumed galaxy-halo SHMR; see Appendix \ref{sec:a3}. 

 We verified that the resulting scaling relations constrained by our models are consistent with the direct measurements for the subset of BASS AGN that have stellar mass estimates (Figure \ref{fig:mbh-ms}). We note that for rare luminous quasars (e.g. $>10^{45}~L_{\rm bol}$), the SED measurements for stellar masses may be overestimated but do not bias the average statistics. Our $M_{\rm{BH}}-M_{*}$ relations did not rely on these estimates, however, and provided independent constraints on its parameters. Our best-fit scaling relations are also similar to previous direct measurements using other AGN samples \citep{Reines:2015,Shankar:2016,Suh:2020}, although the slope is shallower (but consistent to within error; see Figure \ref{fig:mbh-ms}). The normalization is similar to those found for local AGN, lying below the relation for nearby inactive ellipticals \citep{Kormendy:2013}.
 
The summary statistics of the data and models are shown in Figures \ref{fig:stats1} and \ref{fig:stats2}. The X-ray luminosity function and cross-correlation function of the mock AGN generated by the best-fit model parameters are fairly well-matched to the BASS DR2 measurements. 
In general, Model 1 more often produced AGN number densities that were systematically higher than the measurements for luminous AGN. This is due to the wider range of halo mass for a given SMBH mass in Model 1 (see fig \ref{fig:mbhdist}); more higher-luminosity AGN reside in less-massive halos in Model 1 than in Model 2, and these smaller halos are more numerous.  
However, despite different assumptions for the relationship between SMBH mass and halo mass, both models produced realizations that reproduced the overall clustering statistics with consistent SMBH$-$galaxy scaling relations. The Bayes factor \citep[e.g.,][]{kass:1995} between the models (i.e., the relative probabilities of the models given the data; $B_{21} = P(M_{2}|D)/P(M_{1}|D)$), estimated by the distributions of the relative acceptance fractions in the ABC algorithm, is $1.30\pm 0.17$. This indicates that Model 2 is more probable than Model 1 by $\sim 1.75\sigma$, given the data.

\subsection{Halo Properties of Mock SMBHs and AGN}
The distribution of black hole masses as a function of peak (sub)halo mass is shown in Figure \ref{fig:mbhdist} for each best-fit model. These distributions arise from the \Mbh$-M_{*}$ and $M_{*}-M_{peak}$ relations (and their associated scatters), as well as the additional correlation (or not) between \Mbh\ and $M_{peak}$ at fixed $M_{*}$. The median relations are different between the two models, as is the scatter. Model 1 produces a much wider range of black hole masses in a given bin of halo mass than Model 2. These differences result in different distributions of halo masses for the AGN mocks generated by each model despite each producing similar clustering and abundance statistics. This suggests that there are degeneracies between the models that the overall clustering and abundance of AGN alone cannot cleanly break. Table \ref{tab:tab} lists the halo properties for mock AGN (with BASS AGN luminosities) generated by each model. 
The median parent halo masses of the mock AGN in Model 1 is $12.0\pm0.2$ dex [\Msun], while it is $12.1\pm0.2$ dex [\Msun] in Model 2. The average halo masses are $\sim 13.3$ dex [\Msun] for in both models.

The AGN satellite fractions (associated with the BASS AGN selection) are similar between each model: $0.18\pm0.01$ and $0.17\pm0.01$ for Models 1 and 2, respectively. These fractions agree with several other studies that have constrained the X-ray AGN satellite fraction via other methods \citep{Allevato:2012,Leauthaud:2015}. Both our models assumed the same scaling relations and accretion probabilities between SMBHs in centrals and satellite galaxies. Therefore, these satellite fractions arise purely due to the AGN selection function, without any assumed differences or dependencies on where the galaxy/SMBH resides in the parent halo.

\subsection{Implementing a 3rd Constraint}
\label{sec:third}
We repeated the analysis using an independent measurement for the $M_{\rm BH}-M_{halo}$ relationship (from \citealt{Marasco:2021}) as a third constraint. We found that the best-fit \Mbh$-M_{*}$ relation in this analysis for Model 1 has a normalization of $8.10^{+0.09}_{-0.07}$ dex [\Msun], an intrinsic scatter of $0.12^{+0.11}_{-0.09}$ dex, and slope of $0.92^{+0.14}_{-0.12}$. For Model 2, the best-fit parameters of the \Mbh$-M_{*}$ relation has a normalization of $7.91^{+0.14}_{-0.17}$, an intrinsic scatter of $0.35^{+0.15}_{-0.18}$, and a slope of $0.93^{+0.16}_{-0.17}$. While these values were consistent within the uncertainties with the previous analysis, a larger slope and smaller scatter were preferred. 

Most notably, the Bayes Factor between the two models increased to $4.1\pm 0.5$ when including this third constraint, calculated from their relative acceptance fractions. Model 2 became more probable than Model 1 by $4.6\sigma$.

\begin{figure*}
\centering
\includegraphics[width=.7\textwidth]{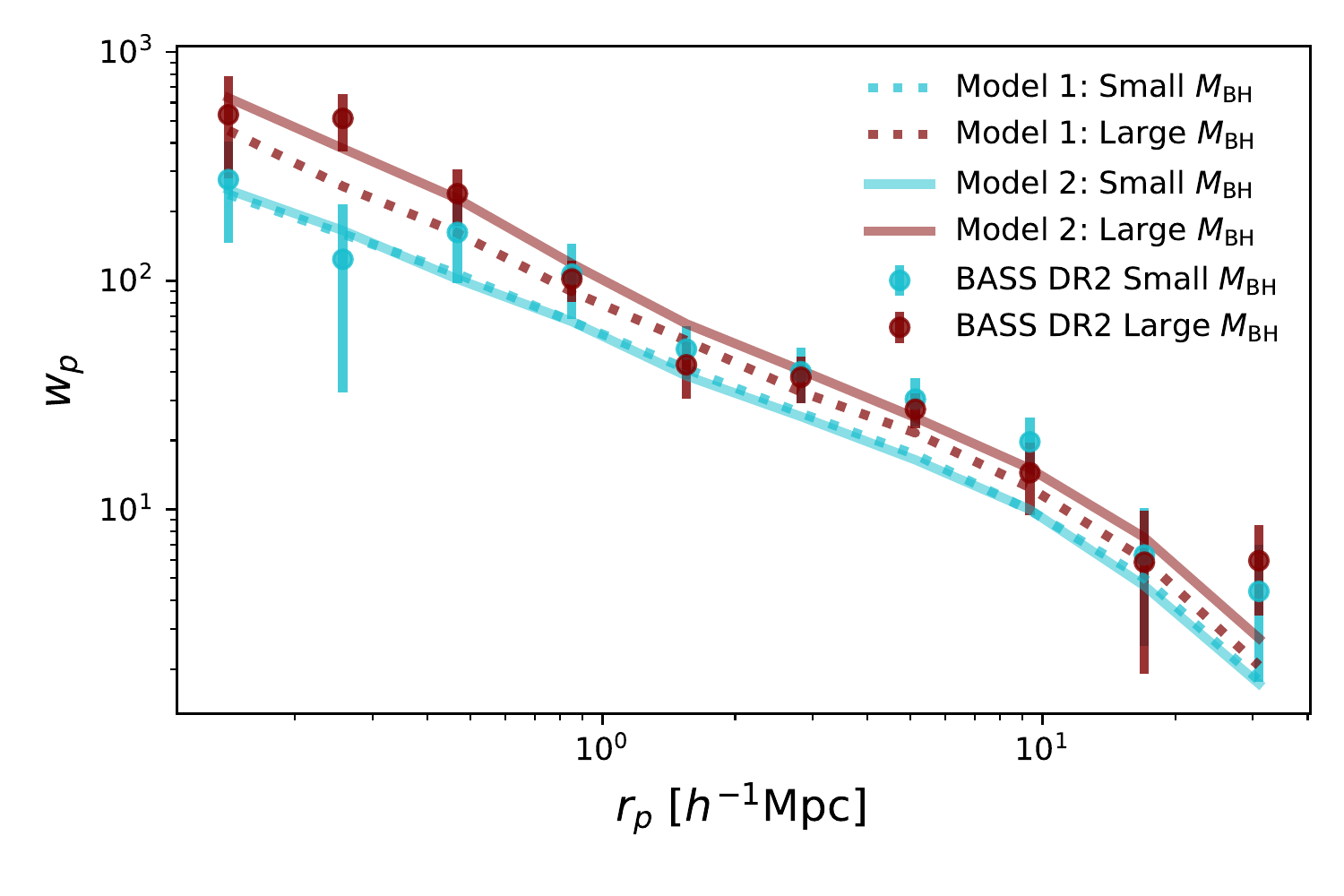}
\caption{Projected correlation functions of the large black hole mass bin (maroon data points) and small black hole mass bin (cyan data points), compared to the two model predictions (Model 1, dotted lines; Model 2, solid lines). Both models show clustering differences between each bin on all scales; however, the differences are greater for Model 2, especially on scales of the 1-halo term ($<1$ \Mpch). This model, in which SMBH mass tightly correlates with halo mass, better matches the data measurements on 1-halo scales.} 
\label{fig:wpmbh}
\end{figure*}

\begin{table}
    \centering
\begin{tabular}{ | c | c  c  | }
\hline
Mock BASS AGN properties  & {\bf Model 1} & {\bf Model 2}  \\ 
 \hline
Avg. host $M_{vir}$ [dex~\Msun]& $13.27\pm0.06$ & $13.34\pm0.08$ \\  
Med. host $M_{vir}$ [dex~\Msun]& 1$2.00\pm 0.16$ & $12.10\pm0.18$ \\
AGN satellite fraction & $0.19\pm0.01$ & $0.18\pm0.01$\\
\hline
\end{tabular}
    \caption{Halo properties of mock AGN generated by Model 1 vs. Model 2, including the average and median virial masses of the parent host halos and the AGN satellite fraction. The mock AGN are chosen with the same luminosity distribution as the BASS DR2 AGN sample.}
    \label{tab:tab}
\end{table}

\subsection{Clustering Trends with $M_{BH}$}
\label{sec:mbh}

To further break degeneracies between the two AGN-halo models, we investigated the clustering trends with black hole mass predicted by each, and compared those trends to the BASS survey measurements. 

Using our two defined bins of black hole mass (Fig. \ref{fig:mbh_vs_z}), we measured the cross-correlation function of each AGN subsample. We then selected mock AGN from each model generation with the same distribution of black hole masses as the \Mbh\ bins, and computed their cross-correlation functions. 
The resulting model correlation functions were calculated by averaging 40 mock generations populated in the Unitsim halo catalog.

The resulting correlation functions of the small and large \Mbh\ bins as measured by the data, Model 1, and Model 2 are shown in Figure \ref{fig:wpmbh}. The data show mild differences between each bin; the AGN with more massive black holes are more clustered on scales of the 1-halo term ($<1$ \Mpch) than AGN with less massive black holes. This was found to be the case independent of \Mbh\ measurement method (appendex \ref{sec:a1}).
However, the current statistics limit a very significant difference.

The two models show differences between the large and small black hole mass bins on all scales, but especially on small scales. Model 2 predicts larger clustering amplitudes for the massive bin. The simpler Model 1 predicts more mild clustering differences between the two \Mbh\ bins, where the more massive mock AGN are less clustered than what is predicted in Model 2. We find that Model 2 is slightly more consistent with the measurements from the data, although better statistics are needed to firmly distinguish between the two model assumptions. 
The reduced correlated $\chi^{2}$ values of the massive black hole bin (where Model 1 and Model 2 are distinct) are 1.3 and 1.1 for Model 1 and 2, respectively.

\section{Discussion}
\label{sec:dis}

In this study, we tested two relatively simple models for how SMBHs and AGN occupy their host dark matter halos. We found that by assuming straightforward relationships between the SMBH, its galaxy, and its host dark matter (sub)halo, as well as a universal Eddington Ratio Distribution function, the AGN space densities and clustering on both 1- and 2-halo scales can be reproduced. We assumed no dependencies on environment for fueling AGN activity, indicating no evidence for larger-scale ($>100$ $h^{-1}$kpc) mechanisms triggering the majority of local, moderate-luminosity AGN.
Interestingly, a larger fraction of close companions \citep{Koss:2010}, dual AGN \citep[e.g.,][]{Koss:2011:L42,Koss:2012:L22,Koss:2016}, and ’hidden' nuclear mergers ($<3$ kpc, \citealt{Koss:2018:214a}) have been found in the BASS sample; however, we did not see a preference for overdense environments on the larger scales probed here.

We also note that while the SMBH--galaxy relations in our simple models only rely on stellar mass, there are other galaxy parameters that have been linked to BASS AGN. Compared to inactive galaxies of similar mass, BASS AGN are more often hosted by gas-rich spiral galaxies \citep{Koss:2011} with higher star formation rates \citep{Mushotzky:2014:L34,Shimizu:2017,Ichikawa:2019} and more molecular gas \citep{Koss:2021}.  Future studies will add more sophistication to improve the AGN--galaxy connection implemented in these models. However, we emphasize that our main goal here was to see whether the host galaxy {\it and} the host halo influence \Mbh\ and the distribution of local AGN rather than the host galaxy alone, as indicated by the scatter in the $M_{\rm BH}-M_{\rm halo}$ relationship. Since stellar mass is the primary galaxy parameter correlated to $M_{\rm halo}$ (and is also correlated with \Mbh), it is the first-order galaxy parameter to control for and the main galaxy parameter we connected to \Mbh. 

Both AGN-halo models tested in this work naturally predicted broad distributions of AGN host halo mass due to the broad distribution of accretion rates.
Previously, typical halo masses of AGN samples have been calculated from the 2-halo term clustering amplitudes. This assumes, however, that the AGN reside in a narrow range of halo mass, which is most likely not the case given recent ERDF constraints \citep{Georgakakis:2019,Jones:2019,Aird:2021}. Our results show that the intrinsic scatter between AGN and halo properties is necessary to understand before the full characterization of the host halo masses can be inferred from AGN clustering measurements. Conversely, this implies that the 'typical' halo masses traditionally calculated by 2-halo term clustering amplitudes do not necessarily correspond to the average or median halo masses of the AGN sample. Previous estimates of the typical halo mass from {\it Swift}/BAT AGN bias measurements have found $\log(M_{halo})\sim 12.8$ \Msun \citep{Cappelluti:2010,Krumpe:2017,Powell:2018}, which is in between the median ($\sim 12$ dex) and average ($\sim 13.3$ dex) halo masses of the mock AGN generated by both our models. Similar conclusions regarding discrepancies between the bias-derived 'typical' and median AGN halo masses were drawn in \citealt{Aird:2021}, in which mock AGN were populated in simulated galaxies based on galaxy properties. The importance of scatter in the scaling relations (and satellite fraction) on interpreting X-ray AGN clustering has also been recently investigated at $z=1.2$ using similar forward-modeling methods, showing that the AGN bias at moderate redshifts is highly dependent on these parameters \citep{Viitanen:2021}.

Our analysis also showed that clustering trends with black hole mass provide additional tests for how strongly SMBHs are correlated with their (sub)halos; Model 2, which assumed a monotonic relationship between SMBH mass and (sub)halo mass for fixed stellar mass, predicted larger clustering differences between AGN with large and small SMBH masses. We found that Model 2 was marginally better-matched to the data measurements. The clustering trends with \Mbh\ predicted by the models are also qualitatively consistent with results found at slightly higher redshifts.
\citealt{Krumpe:2015} measured stronger clustering on 2-halo scales for X-ray AGN at $z\sim 0.3$ with more massive black holes than AGN with less massive black holes. Additionally, they found no trends with Eddington ratio while controlling for black hole mass, which is also expected in both of our models.

The Bayes factor between our two models, as well as the observed clustering trends with black hole mass, demonstrated that Model 2 is preferred by the BASS DR2 dataset over Model 1. This would indicate that the total dark matter mass of the galaxy, rather than baryonic mass, is fundamentally connected to SMBH formation and growth. Several independent measurements of the local black hole mass$-$halo mass relation support this idea, as there have been claims of strong correlations between SMBH mass and various proxies for host halo mass  \citep{Ferrarese:2002,Bandara:2009,Lakhchaura:2019,Smith:2021,Robinson:2021}.
Other work has claimed that these correlations only hold for bulge-dominated galaxies \citep[e.g.,][]{Kormendy:2013,Sabra:2015} and arise as a result of the coevolution between the black hole and the spheroidal galaxy component of the galaxy due to e.g., mergers. However, the recent, updated study of
\citealt{Marasco:2021} investigated 55 nearby galaxies with dynamically measured black hole masses and host halo mass estimates inferred by globular cluster dynamics or spatially-resolved rotation curves, and found that the \Mbh$-M_{h}$ correlation holds for both early and late type systems. When using this result as a third constraint, Model 2 was significantly preferred over Model 1.

\begin{figure}
    \centering
    \includegraphics[width=.49\textwidth]{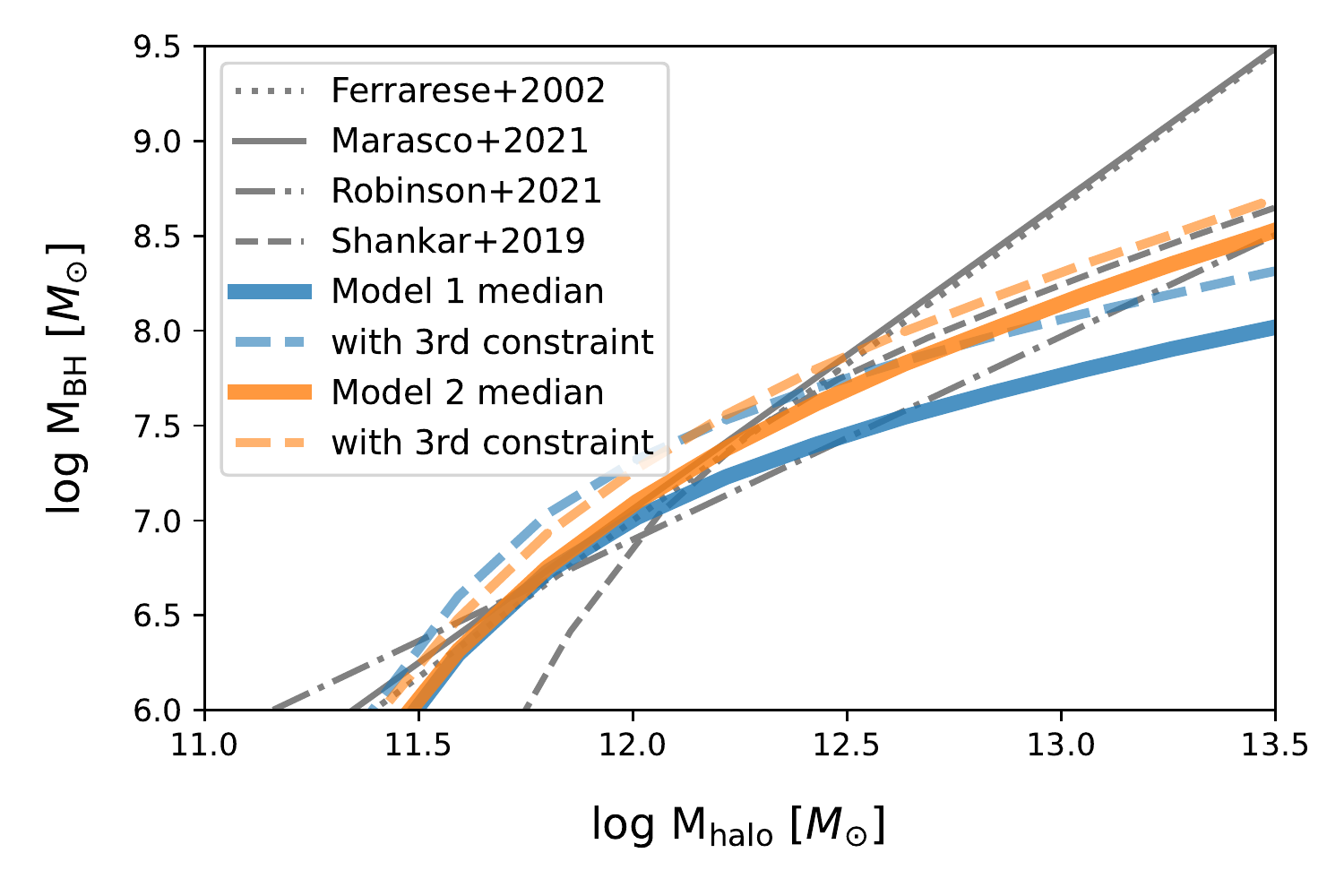}
    \caption{ Median SMBH mass-halo mass relations predicted by each model (Model 1, blue; Model 2, orange) compared to those reported in the literature that were measured using individual galaxies and various proxies for halo mass (gray lines). The dashed lines correspond to the relations predicted by the best fit models when imposing the 3rd constraint.}
    \label{fig:mbhmhlit}
\end{figure}

Figure \ref{fig:mbhmhlit} shows the average \Mbh$-M_{halo}$ relations for both of our models compared to several previous observational measurements. Both models tend to lie below the measurements at high masses, which may be due to the range of black hole masses in our data sample; we have very few AGN with \Mbh\ $>10^{9}~M_{\odot}$ ($11$; $\sim 2\%$). Expanded samples with a larger range of \Mbh\ would better test the high-mass halo regime  $>10^{13}$ \Msun. However,
the slope and scatter of \Mbh$-M_{peak}$ associated with Model 2 are much more consistent with these recent observational measurements \citep[e.g.,][]{Marasco:2021,Robinson:2021} than Model 1. There is especially good agreement with the high mass end of the relation derived by \citealt{Shankar:2020}, which used 2-halo amplitudes from several AGN clustering measurements and abundance arguments to constrain \Mbh$-M_{halo}$. 

\subsection{Interpretations of a tight correlation between SMBH mass and halo mass}
Despite the vastly different scales between SMBHs and their host halos, there are physical motivations for a correlation between them. Several hydrodynamic and semi-analytic models have suggested that the binding energy of dark matter halos, rather than total stellar mass, determines the masses of their central supermassive black holes \citep[e.g.,][]{Booth:2010,Bower:2017,Marasco:2021}. In these scenarios, baryons and dark matter accrete onto halos. The gas cools to eventually provide fuel for star formation and AGN activity. Feedback from star formation regulates both stellar and black hole growth in halos with masses $\lesssim10^{12}$\Msun, while more massive halos prevent outflows from efficiently limiting cold gas at the centers of galaxies due to their hot coronas. This causes rapid black hole growth in the massive systems until the total energy output of AGN feedback exceeds a threshold associated with the binding energy of the halo \citep[e.g.,][]{Bower:2017,Chen:2020}. After this point, the gas is unable to cool and provide sufficient fuel for star formation or SMBH growth, star formation quenches, and 'maintenance-mode' AGN feedback \citep[e.g.,][]{Croton:2006} regulates further significant SMBH accretion. The connection between black hole accretion and the halo's binding energy results in a tight correlation between black hole mass and halo mass in systems $\gtrsim 10^{12}$\Msun. 
While the $M_{\rm{BH}}-M_{halo}$ relations in the Eagle and IllustrusTNG simulations depend on seeding prescription and resolution at the low mass end, the slopes of the relations $> 10^{12}$\Msun\ (as presented in \citealt{Bower:2017} and \citealt{Weinberger:2018}, respectively) match that of Model 2. Probing the extreme mass scales (low and high mass systems) will better test this scenario.

A correlation between supermassive black hole mass and halo mass could alternatively imply that the most massive SMBHs in the local universe preferably reside in the densest regions of the universe due to their formation. The most massive halos trace the highest overdensities; if SMBHs in the earlier universe formed via many gas–rich mergers, or from
SMBH seeds in massive metal–poor gas halos \citep[e.g.,][]{Wise:2019,Inayoshi:2020}, then it would mean that the oldest-formed, massive SMBHs today should statistically lie in massive halos. However, there have been inconclusive studies of high-z quasars and whether they preferentially reside in overdensities or not  \citep[e.g.,][and references therein]{Husband:2013,Mazzucchelli:2017,Meyer:2022}. Measuring the environments of quasars at moderate-to-high redshifts with larger samples (with future surveys via e.g., eROSITA and JWST) will better test the conditions of SMBH formation.

Lastly, there has been recent interest in the hypothesis that dark matter is composed of primordial black holes, originally suggested by \citealt{Hawking:1971} and rejuvenated by the LIGO and Virgo gravitational wave discoveries. In one such theory, black holes were produced in the early universe during QCD phase transitions \citep[e.g.,][]{Hasinger:2020,Carr:2021,Cappelluti:2022}, forming a broad mass function that is not currently ruled out by observational constraints. At high redshifts, smaller black holes would cluster around larger black holes while growing via Bondi accretion and mergers. Eventually, halos would form whose mass is proportional to the central SMBH in the following way: $M_{halo}=\frac{f_{\rm dm}(M_{\rm BH})}{M_{\rm BH}}$, where $f_{\rm dm}$ refers to the fraction of dark matter in primordial black holes of a given mass \citep{Cappelluti:2022}. After baryons condense to form galaxies, they would grow alongside the SMBH and halo, but the \Mbh$-M_{halo}$ correlation would be maintained. Based on the shape of the assumed mass spectrum in \citealt{Cappelluti:2022} (which is extrapolated for \Mbh$>10^{6}$ M$_{\odot}$), this correlation is predicted to have a slope of $\sim 0.6$ (for $\log M_{\rm BH}>7.5$ M$_{\odot}$), similar to our models. 
This theory will be further tested by JWST and eventually LISA.

Characterizing the local \Mbh$-M_{halo}$ relation would provide a benchmark for black hole-galaxy coevolutionary scenarios implemented in hydrodynamic and semi-analytic simulations. Future, deeper datasets will tighten the constraints on the relationships between SMBHs, galaxies, and their dark matter halos, and will further test SMBH formation and assembly.

\section{Summary}
\label{sec:sum}

We have used the clustering and space densities of AGN from the BASS DR2 sample to investigate the local relationship between supermassive black hole mass and host (sub)halo mass.
By populating AGN activity into cosmological simulations and assuming a universal Eddington Ratio Distribution function, two model assumptions were tested: (1) SMBH mass primarily depends on its host galaxy mass, and (2) there is a secondary correlation between SMBH mass and host peak (sub)halo mass. Our main conclusions are as follows:

\begin{itemize}
    \item Both simple AGN-halo models were able to reproduce the measured AGN luminosity function and overall correlation function, with consistent constraints on the scaling relation between black hole mass and stellar mass and a universal Eddington Ratio Distribution Function. This points to stochastic, in-situ fueling, rather than large-scale environmental mechanisms triggering local AGN.
    
   \item The distributions of mock AGN host halo masses were different for each model, despite being able to generate similar AGN clustering statistics and space densities. We show that clustering trends with black hole mass can distinguish between them.

   \item BASS AGN with large black hole masses were found to be more clustered on 1-halo scales than AGN with small black hole masses. The halo model in which black hole mass is more tightly correlated with (sub)halo mass does a better job at reproducing these differences. 
   
   \item The Bayes factor between the two models, as well as the observed clustering trends with black hole mass, indicated that the data preferred the scenario in which SMBH mass is correlated with host dark matter (sub)halo mass for fixed stellar mass (by $\sim 2-5\sigma$, depending on the constraints used) over a model absent of this correlation. This may indicate that the total dark matter mass is connected to SMBH formation and growth.

\end{itemize}

Future spectroscopic surveys like eROSITA/4MOST, DESI, and WEAVE-LOFAR will detect and characterize orders of magnitude more AGN than present surveys. This will enable more precise clustering measurements as a function of SMBH and AGN parameters. The models presented in this work will be firmly tested and constrained as a function of redshift, which will improve the understanding of how the assembly of local SMBHs and galaxies is established over cosmic time.\\

\acknowledgments
We thank the anonymous reviewer for helpful comments that improved the paper. MP thanks Risa Weschler for helpful discussions.
We acknowledge support from NASA through ADAP award NNH16CT03C (M.K.) and NASA-Swift 80NSSC18K0505 (N.C.); 
the Israel Science Foundation through grant number 1849/19 (B.T.); 
the European Research Council (ERC) under the European Union's Horizon 2020 research and innovation program, through grant agreement number 950533 (B.T.);
ANID grants CATA-Basal 
FB210003 (C.R.); 
Fondecyt Iniciacion grant 11190831 (C.R.); 
the National Research Foundation of Korea grant NRF-2020R1C1C1005462 and the Japan Society for the Promotion of Science ID: 17321 (K.O.); 
We acknowledge the work done by the 50+ BASS scientists and Swift-BAT team to make this project possible.
The authors gratefully acknowledge the Gauss Centre for Supercomputing e.V. (www.gauss-centre.eu) and the Partnership for Advanced Supercomputing in Europe (PRACE, www.prace-ri.eu) for funding the MultiDark simulation project by providing computing time on the GCS Supercomputer SuperMUC at Leibniz Supercomputing Centre (LRZ, www.lrz.de).
The CosmoSim database used in this paper is a service by the Leibniz-Institute for Astrophysics Potsdam (AIP).
The MultiDark database was developed in cooperation with the Spanish MultiDark Consolider Project CSD2009-00064.\\

\software{AbundanceMatching (\url{https://github.com/yymao/abundancematching}), halotools \citep{Hearin:2017}, astropy \citep{Astropy:2013,astropy:2018}, abcpmc \citep{abcpmc:2015}, matplotlib \citep{Hunter:2007}}\\

\bibliography{references}{}

\begin{thebibliography}{}
\expandafter\ifx\csname natexlab\endcsname\relax\def\natexlab#1{#1}\fi

\bibitem[{{Aird} \& {Coil}(2021)}]{Aird:2021}
{Aird}, J., \& {Coil}, A.~L. 2021, \mnras, 502, 5962

\bibitem[{{Aird} {et~al.}(2018){Aird}, {Coil}, \& {Georgakakis}}]{Aird:2018}
{Aird}, J., {Coil}, A.~L., \& {Georgakakis}, A. 2018, \mnras, 474, 1225

\bibitem[{{Aird} {et~al.}(2022){Aird}, {Coil}, \& {Kocevski}}]{Aird:2022}
{Aird}, J., {Coil}, A.~L., \& {Kocevski}, D.~D. 2022, arXiv e-prints,
  arXiv:2201.11756

\bibitem[{Akeret {et~al.}(2015)Akeret, Refregier, Amara, Seehars, \&
  Hasner}]{abcpmc:2015}
Akeret, J., Refregier, A., Amara, A., Seehars, S., \& Hasner, C. 2015, Journal
  of Cosmology and Astroparticle Physics, 2015, 043

\bibitem[{{Allevato} {et~al.}(2021){Allevato}, {Shankar}, {Marsden}, {Rasulov},
  {Viitanen}, {Georgakakis}, {Ferrara}, \& {Finoguenov}}]{Allevato:2021}
{Allevato}, V., {Shankar}, F., {Marsden}, C., {et~al.} 2021, \apj, 916, 34

\bibitem[{{Allevato} {et~al.}(2011){Allevato}, {Finoguenov}, {Cappelluti},
  {Miyaji}, {Hasinger}, {Salvato}, {Brusa}, {Gilli}, {Zamorani}, {Shankar},
  {James}, {McCracken}, {Bongiorno}, {Merloni}, {Peacock}, {Silverman}, \&
  {Comastri}}]{Allevato:2011}
{Allevato}, V., {Finoguenov}, A., {Cappelluti}, N., {et~al.} 2011, \apj, 736,
  99

\bibitem[{{Allevato} {et~al.}(2012){Allevato}, {Finoguenov}, {Hasinger},
  {Miyaji}, {Cappelluti}, {Salvato}, {Zamorani}, {Gilli}, {George}, {Tanaka},
  {Brusa}, {Silverman}, {Civano}, {Elvis}, \& {Shankar}}]{Allevato:2012}
{Allevato}, V., {Finoguenov}, A., {Hasinger}, G., {et~al.} 2012, \apj, 758, 47

\bibitem[{{Allevato} {et~al.}(2014){Allevato}, {Finoguenov}, {Civano},
  {Cappelluti}, {Shankar}, {Miyaji}, {Hasinger}, {Gilli}, {Zamorani},
  {Lanzuisi}, {Salvato}, {Elvis}, {Comastri}, \& {Silverman}}]{Allevato:2014}
{Allevato}, V., {Finoguenov}, A., {Civano}, F., {et~al.} 2014, \apj, 796, 4

\bibitem[{{Ananna} {et~al.}(2022){Ananna}, {Weigel}, {Trakhtenbrot}, {Koss},
  {Urry}, {Ricci}, {Hickox}, {Treister}, {Bauer}, {Ueda}, {Mushotzky}, {Ricci},
  {Oh}, {Mej{\'\i}a-Restrepo}, {Brok}, {Stern}, {Powell}, {Caglar}, {Ichikawa},
  {Wong}, {Harrison}, \& {Schawinski}}]{Ananna:2022}
{Ananna}, T.~T., {Weigel}, A.~K., {Trakhtenbrot}, B., {et~al.} 2022, \apjs,
  261, 9

\bibitem[{{Astropy Collaboration} {et~al.}(2013){Astropy Collaboration},
  {Robitaille}, {Tollerud}, {Greenfield}, {Droettboom}, {Bray}, {Aldcroft},
  {Davis}, {Ginsburg}, {Price-Whelan}, {Kerzendorf}, {Conley}, {Crighton},
  {Barbary}, {Muna}, {Ferguson}, {Grollier}, {Parikh}, {Nair}, {Unther},
  {Deil}, {Woillez}, {Conseil}, {Kramer}, {Turner}, {Singer}, {Fox}, {Weaver},
  {Zabalza}, {Edwards}, {Azalee Bostroem}, {Burke}, {Casey}, {Crawford},
  {Dencheva}, {Ely}, {Jenness}, {Labrie}, {Lim}, {Pierfederici}, {Pontzen},
  {Ptak}, {Refsdal}, {Servillat}, \& {Streicher}}]{Astropy:2013}
{Astropy Collaboration}, {Robitaille}, T.~P., {Tollerud}, E.~J., {et~al.} 2013,
  \aap, 558, A33

\bibitem[{{Astropy Collaboration} {et~al.}(2018){Astropy Collaboration},
  {Price-Whelan}, {Sip{\H{o}}cz}, {G{\"u}nther}, {Lim}, {Crawford}, {Conseil},
  {Shupe}, {Craig}, {Dencheva}, {Ginsburg}, {Vand erPlas}, {Bradley},
  {P{\'e}rez-Su{\'a}rez}, {de Val-Borro}, {Aldcroft}, {Cruz}, {Robitaille},
  {Tollerud}, {Ardelean}, {Babej}, {Bach}, {Bachetti}, {Bakanov}, {Bamford},
  {Barentsen}, {Barmby}, {Baumbach}, {Berry}, {Biscani}, {Boquien}, {Bostroem},
  {Bouma}, {Brammer}, {Bray}, {Breytenbach}, {Buddelmeijer}, {Burke},
  {Calderone}, {Cano Rodr{\'\i}guez}, {Cara}, {Cardoso}, {Cheedella}, {Copin},
  {Corrales}, {Crichton}, {D'Avella}, {Deil}, {Depagne}, {Dietrich}, {Donath},
  {Droettboom}, {Earl}, {Erben}, {Fabbro}, {Ferreira}, {Finethy}, {Fox},
  {Garrison}, {Gibbons}, {Goldstein}, {Gommers}, {Greco}, {Greenfield},
  {Groener}, {Grollier}, {Hagen}, {Hirst}, {Homeier}, {Horton}, {Hosseinzadeh},
  {Hu}, {Hunkeler}, {Ivezi{\'c}}, {Jain}, {Jenness}, {Kanarek}, {Kendrew},
  {Kern}, {Kerzendorf}, {Khvalko}, {King}, {Kirkby}, {Kulkarni}, {Kumar},
  {Lee}, {Lenz}, {Littlefair}, {Ma}, {Macleod}, {Mastropietro}, {McCully},
  {Montagnac}, {Morris}, {Mueller}, {Mumford}, {Muna}, {Murphy}, {Nelson},
  {Nguyen}, {Ninan}, {N{\"o}the}, {Ogaz}, {Oh}, {Parejko}, {Parley}, {Pascual},
  {Patil}, {Patil}, {Plunkett}, {Prochaska}, {Rastogi}, {Reddy Janga},
  {Sabater}, {Sakurikar}, {Seifert}, {Sherbert}, {Sherwood-Taylor}, {Shih},
  {Sick}, {Silbiger}, {Singanamalla}, {Singer}, {Sladen}, {Sooley},
  {Sornarajah}, {Streicher}, {Teuben}, {Thomas}, {Tremblay}, {Turner},
  {Terr{\'o}n}, {van Kerkwijk}, {de la Vega}, {Watkins}, {Weaver}, {Whitmore},
  {Woillez}, {Zabalza}, \& {Astropy Contributors}}]{astropy:2018}
{Astropy Collaboration}, {Price-Whelan}, A.~M., {Sip{\H{o}}cz}, B.~M., {et~al.}
  2018, \aj, 156, 123

\bibitem[{{Azadi} {et~al.}(2017){Azadi}, {Coil}, {Aird}, {Reddy}, {Shapley},
  {Freeman}, {Kriek}, {Leung}, {Mobasher}, {Price}, {Sanders}, {Shivaei}, \&
  {Siana}}]{azadi:2017}
{Azadi}, M., {Coil}, A.~L., {Aird}, J., {et~al.} 2017, \apj, 835, 27

\bibitem[{{Bandara} {et~al.}(2009){Bandara}, {Crampton}, \&
  {Simard}}]{Bandara:2009}
{Bandara}, K., {Crampton}, D., \& {Simard}, L. 2009, \apj, 704, 1135

\bibitem[{{Baumgartner} {et~al.}(2013){Baumgartner}, {Tueller}, {Markwardt},
  {Skinner}, {Barthelmy}, {Mushotzky}, {Evans}, \&
  {Gehrels}}]{Baumgartner:2013}
{Baumgartner}, W.~H., {Tueller}, J., {Markwardt}, C.~B., {et~al.} 2013, \apjs,
  207, 19

\bibitem[{{Behroozi} {et~al.}(2010){Behroozi}, {Conroy}, \&
  {Wechsler}}]{Behroozi:2010}
{Behroozi}, P.~S., {Conroy}, C., \& {Wechsler}, R.~H. 2010, \apj, 717, 379

\bibitem[{{Behroozi} {et~al.}(2014){Behroozi}, {Wechsler}, {Lu}, {Hahn},
  {Busha}, {Klypin}, \& {Primack}}]{Behroozi:2014}
{Behroozi}, P.~S., {Wechsler}, R.~H., {Lu}, Y., {et~al.} 2014, \apj, 787, 156

\bibitem[{{Behroozi} {et~al.}(2013){Behroozi}, {Wechsler}, \& {Wu}}]{rockstar}
{Behroozi}, P.~S., {Wechsler}, R.~H., \& {Wu}, H.-Y. 2013, \apj, 762, 109

\bibitem[{{Bonne} {et~al.}(2015){Bonne}, {Brown}, {Jones}, \&
  {Pimbblet}}]{Bonne:2015}
{Bonne}, N.~J., {Brown}, M. J.~I., {Jones}, H., \& {Pimbblet}, K.~A. 2015,
  \apj, 799, 160

\bibitem[{{Booth} \& {Schaye}(2010)}]{Booth:2010}
{Booth}, C.~M., \& {Schaye}, J. 2010, \mnras, 405, L1

\bibitem[{{Bower} {et~al.}(2017){Bower}, {Schaye}, {Frenk}, {Theuns},
  {Schaller}, {Crain}, \& {McAlpine}}]{Bower:2017}
{Bower}, R.~G., {Schaye}, J., {Frenk}, C.~S., {et~al.} 2017, \mnras, 465, 32

\bibitem[{{Caglar} {et~al.}(2020){Caglar}, {Burtscher}, {Brandl}, {Brinchmann},
  {Davies}, {Hicks}, {Koss}, {Lin}, {Maciejewski}, {M{\"u}ller-S{\'a}nchez},
  {Riffel}, {Riffel}, {Rosario}, {Schartmann}, {Schnorr-M{\"u}ller}, {Shimizu},
  {Storchi-Bergmann}, {Veilleux}, {Orban de Xivry}, \& {Bennert}}]{Caglar:2020}
{Caglar}, T., {Burtscher}, L., {Brandl}, B., {et~al.} 2020, \aap, 634, A114

\bibitem[{{Cappelluti} {et~al.}(2010){Cappelluti}, {Ajello}, {Burlon},
  {Krumpe}, {Miyaji}, {Bonoli}, \& {Greiner}}]{Cappelluti:2010}
{Cappelluti}, N., {Ajello}, M., {Burlon}, D., {et~al.} 2010, \apjl, 716, L209

\bibitem[{{Cappelluti} {et~al.}(2012){Cappelluti}, {Allevato}, \&
  {Finoguenov}}]{Cappelluti:2012}
{Cappelluti}, N., {Allevato}, V., \& {Finoguenov}, A. 2012, Advances in
  Astronomy, 2012, 853701

\bibitem[{{Cappelluti} {et~al.}(2022){Cappelluti}, {Hasinger}, \&
  {Natarajan}}]{Cappelluti:2022}
{Cappelluti}, N., {Hasinger}, G., \& {Natarajan}, P. 2022, \apj, 926, 205

\bibitem[{{Carr} {et~al.}(2021){Carr}, {Kohri}, {Sendouda}, \&
  {Yokoyama}}]{Carr:2021}
{Carr}, B., {Kohri}, K., {Sendouda}, Y., \& {Yokoyama}, J. 2021, Reports on
  Progress in Physics, 84, 116902

\bibitem[{{Chen} {et~al.}(2020){Chen}, {Faber}, {Koo}, {Somerville}, {Primack},
  {Dekel}, {Rodr{\'\i}guez-Puebla}, {Guo}, {Barro}, {Kocevski}, {van der Wel},
  {Woo}, {Bell}, {Fang}, {Ferguson}, {Giavalisco}, {Huertas-Company}, {Jiang},
  {Kassin}, {Lin}, {Liu}, {Luo}, {Luo}, {Pacifici}, {Pandya}, {Salim}, {Shu},
  {Tacchella}, {Terrazas}, \& {Yesuf}}]{Chen:2020}
{Chen}, Z., {Faber}, S.~M., {Koo}, D.~C., {et~al.} 2020, \apj, 897, 102

\bibitem[{{Chuang} {et~al.}(2019){Chuang}, {Yepes}, {Kitaura},
  {Pellejero-Ibanez}, {Rodr{\'\i}guez-Torres}, {Feng}, {Metcalf}, {Wechsler},
  {Zhao}, {To}, {Alam}, {Banerjee}, {DeRose}, {Giocoli}, {Knebe}, \&
  {Reyes}}]{Chuang:2019}
{Chuang}, C.-H., {Yepes}, G., {Kitaura}, F.-S., {et~al.} 2019, \mnras, 487, 48

\bibitem[{{Coil} {et~al.}(2009){Coil}, {Georgakakis}, {Newman}, {Cooper},
  {Croton}, {Davis}, {Koo}, {Laird}, {Nandra}, {Weiner}, {Willmer}, \&
  {Yan}}]{Coil:2009}
{Coil}, A.~L., {Georgakakis}, A., {Newman}, J.~A., {et~al.} 2009, \apj, 701,
  1484

\bibitem[{{Comparat} {et~al.}(2019){Comparat}, {Merloni}, {Salvato}, {Nandra},
  {Boller}, {Georgakakis}, {Finoguenov}, {Dwelly}, {Buchner}, {Del Moro},
  {Clerc}, {Wang}, {Zhao}, {Prada}, {Yepes}, {Brusa}, {Krumpe}, \&
  {Liu}}]{Comparat:2019}
{Comparat}, J., {Merloni}, A., {Salvato}, M., {et~al.} 2019, \mnras, 487, 2005

\bibitem[{{Conroy} {et~al.}(2006){Conroy}, {Wechsler}, \&
  {Kravtsov}}]{Conroy:2006}
{Conroy}, C., {Wechsler}, R.~H., \& {Kravtsov}, A.~V. 2006, \apj, 647, 201

\bibitem[{{Croton} {et~al.}(2006){Croton}, {Springel}, {White}, {De Lucia},
  {Frenk}, {Gao}, {Jenkins}, {Kauffmann}, {Navarro}, \&
  {Yoshida}}]{Croton:2006}
{Croton}, D.~J., {Springel}, V., {White}, S. D.~M., {et~al.} 2006, \mnras, 365,
  11

\bibitem[{{DeGraf} \& {Sijacki}(2017)}]{DeGraf:2017}
{DeGraf}, C., \& {Sijacki}, D. 2017, \mnras, 466, 3331

\bibitem[{{Ding} {et~al.}(2022){Ding}, {Silverman}, {Treu}, {Li}, {Bhowmick},
  {Menci}, {Volonteri}, {Blecha}, {Di Matteo}, \& {Dubois}}]{Ding:2022}
{Ding}, X., {Silverman}, J.~D., {Treu}, T., {et~al.} 2022, arXiv e-prints,
  arXiv:2205.04481

\bibitem[{{DiPompeo} {et~al.}(2017){DiPompeo}, {Hickox}, {Eftekharzadeh}, \&
  {Myers}}]{DiPompeo:2017}
{DiPompeo}, M.~A., {Hickox}, R.~C., {Eftekharzadeh}, S., \& {Myers}, A.~D.
  2017, \mnras, 469, 4630

\bibitem[{{Eftekharzadeh} {et~al.}(2015){Eftekharzadeh}, {Myers}, {White},
  {Weinberg}, {Schneider}, {Shen}, {Font-Ribera}, {Ross}, {Paris}, \&
  {Streblyanska}}]{Eftekharzade:2015}
{Eftekharzadeh}, S., {Myers}, A.~D., {White}, M., {et~al.} 2015, \mnras, 453,
  2779

\bibitem[{{Ferrarese}(2002)}]{Ferrarese:2002}
{Ferrarese}, L. 2002, \apj, 578, 90

\bibitem[{{Galloway} {et~al.}(2015){Galloway}, {Willett}, {Fortson},
  {Cardamone}, {Schawinski}, {Cheung}, {Lintott}, {Masters}, {Melvin}, \&
  {Simmons}}]{Galloway:2015}
{Galloway}, M.~A., {Willett}, K.~W., {Fortson}, L.~F., {et~al.} 2015, \mnras,
  448, 3442

\bibitem[{{Georgakakis} {et~al.}(2019){Georgakakis}, {Comparat}, {Merloni},
  {Ciesla}, {Aird}, \& {Finoguenov}}]{Georgakakis:2019}
{Georgakakis}, A., {Comparat}, J., {Merloni}, A., {et~al.} 2019, \mnras, 487,
  275

\bibitem[{{Hahn} {et~al.}(2017){Hahn}, {Vakili}, {Walsh}, {Hearin}, {Hogg}, \&
  {Campbell}}]{Hahn:2017}
{Hahn}, C., {Vakili}, M., {Walsh}, K., {et~al.} 2017, \mnras, 469, 2791

\bibitem[{{Hasinger}(2020)}]{Hasinger:2020}
{Hasinger}, G. 2020, \jcap, 2020, 022

\bibitem[{{Hawking}(1971)}]{Hawking:1971}
{Hawking}, S. 1971, \mnras, 152, 75

\bibitem[{{He} {et~al.}(2018){He}, {Akiyama}, {Bosch}, {Enoki}, {Harikane},
  {Ikeda}, {Kashikawa}, {Kawaguchi}, {Komiyama}, {Lee}, {Matsuoka}, {Miyazaki},
  {Nagao}, {Nagashima}, {Niida}, {Nishizawa}, {Oguri}, {Onoue}, {Oogi},
  {Ouchi}, {Schulze}, {Shirasaki}, {Silverman}, {Tanaka}, {Tanaka}, {Toba},
  {Uchiyama}, \& {Yamashita}}]{He:2018}
{He}, W., {Akiyama}, M., {Bosch}, J., {et~al.} 2018, Publications of the
  Astronomical Society of Japan, 70, S33

\bibitem[{{Hearin} \& {Watson}(2013)}]{Hearin:2013}
{Hearin}, A.~P., \& {Watson}, D.~F. 2013, \mnras, 435, 1313

\bibitem[{{Hearin} {et~al.}(2017){Hearin}, {Campbell}, {Tollerud}, {Behroozi},
  {Diemer}, {Goldbaum}, {Jennings}, {Leauthaud}, {Mao}, {More}, {Parejko},
  {Sinha}, {Sip{\"o}cz}, \& {Zentner}}]{Hearin:2017}
{Hearin}, A.~P., {Campbell}, D., {Tollerud}, E., {et~al.} 2017, \aj, 154, 190

\bibitem[{{Hickox} {et~al.}(2009){Hickox}, {Jones}, {Forman}, {Murray},
  {Kochanek}, {Eisenstein}, {Jannuzi}, {Dey}, {Brown}, {Stern}, {Eisenhardt},
  {Gorjian}, {Brodwin}, {Narayan}, {Cool}, {Kenter}, {Caldwell}, \&
  {Anderson}}]{Hickox:2009}
{Hickox}, R.~C., {Jones}, C., {Forman}, W.~R., {et~al.} 2009, \apj, 696, 891

\bibitem[{{Hickox} {et~al.}(2011){Hickox}, {Myers}, {Brodwin}, {Alexander},
  {Forman}, {Jones}, {Murray}, {Brown}, {Cool}, {Kochanek}, {Dey}, {Jannuzi},
  {Eisenstein}, {Assef}, {Eisenhardt}, {Gorjian}, {Stern}, {Le Floc'h},
  {Caldwell}, {Goulding}, \& {Mullaney}}]{Hickox:2011}
{Hickox}, R.~C., {Myers}, A.~D., {Brodwin}, M., {et~al.} 2011, \apj, 731, 117

\bibitem[{{Hopkins} {et~al.}(2008){Hopkins}, {Hernquist}, {Cox}, \& {Kere{\v
  s}}}]{Hopkins:2008}
{Hopkins}, P.~F., {Hernquist}, L., {Cox}, T.~J., \& {Kere{\v s}}, D. 2008,
  \apjs, 175, 356

\bibitem[{{Huchra} {et~al.}(2012){Huchra}, {Macri}, {Masters}, {Jarrett},
  {Berlind}, {Calkins}, {Crook}, {Cutri}, {Erdo{\v g}du}, {Falco}, {George},
  {Hutcheson}, {Lahav}, {Mader}, {Mink}, {Martimbeau}, {Schneider},
  {Skrutskie}, {Tokarz}, \& {Westover}}]{Huchra:2012}
{Huchra}, J.~P., {Macri}, L.~M., {Masters}, K.~L., {et~al.} 2012, \apjs, 199,
  26

\bibitem[{Hunter(2007)}]{Hunter:2007}
Hunter, J.~D. 2007, Computing in Science \& Engineering, 9, 90

\bibitem[{{Husband} {et~al.}(2013){Husband}, {Bremer}, {Stanway}, {Davies},
  {Lehnert}, \& {Douglas}}]{Husband:2013}
{Husband}, K., {Bremer}, M.~N., {Stanway}, E.~R., {et~al.} 2013, \mnras, 432,
  2869

\bibitem[{{Ichikawa} {et~al.}(2019){Ichikawa}, {Ricci}, {Ueda}, {Bauer},
  {Kawamuro}, {Koss}, {Oh}, {Rosario}, {Shimizu}, {Stalevski}, {Fuller},
  {Packham}, \& {Trakhtenbrot}}]{Ichikawa:2019}
{Ichikawa}, K., {Ricci}, C., {Ueda}, Y., {et~al.} 2019, \apj, 870, 31

\bibitem[{{Inayoshi} {et~al.}(2020){Inayoshi}, {Visbal}, \&
  {Haiman}}]{Inayoshi:2020}
{Inayoshi}, K., {Visbal}, E., \& {Haiman}, Z. 2020, \araa, 58, 27

\bibitem[{{Jiang} {et~al.}(2016){Jiang}, {Wang}, {Mo}, {Dong}, {Wang}, \&
  {Zhou}}]{Jiang:2016}
{Jiang}, N., {Wang}, H., {Mo}, H., {et~al.} 2016, \apj, 832, 111

\bibitem[{{Jones} {et~al.}(2006){Jones}, {Peterson}, {Colless}, \&
  {Saunders}}]{Jones:2006}
{Jones}, D.~H., {Peterson}, B.~A., {Colless}, M., \& {Saunders}, W. 2006,
  \mnras, 369, 25

\bibitem[{{Jones} {et~al.}(2019){Jones}, {Hickox}, {Mutch}, {Croton}, {Ptak},
  \& {DiPompeo}}]{Jones:2019}
{Jones}, M.~L., {Hickox}, R.~C., {Mutch}, S.~J., {et~al.} 2019, \apj, 881, 110

\bibitem[{{Kass} \& {Raftery}(1995)}]{kass:1995}
{Kass}, R.~E., \& {Raftery}, A.~E. 1995, JASA, 90, 773

\bibitem[{{Klypin} {et~al.}(2016){Klypin}, {Yepes}, {Gottl{\"o}ber}, {Prada},
  \& {He{\ss}}}]{SMDPL}
{Klypin}, A., {Yepes}, G., {Gottl{\"o}ber}, S., {Prada}, F., \& {He{\ss}}, S.
  2016, \mnras, 457, 4340

\bibitem[{{Kormendy} \& {Ho}(2013)}]{Kormendy:2013}
{Kormendy}, J., \& {Ho}, L.~C. 2013, \araa, 51, 511

\bibitem[{Koss {et~al.}(2012)Koss, Mushotzky, Treister, Veilleux, Vasudevan, \&
  Trippe}]{Koss:2012:L22}
Koss, M., Mushotzky, R., Treister, E., {et~al.} 2012, \apj, 746, L22

\bibitem[{{Koss} {et~al.}(2010){Koss}, {Mushotzky}, {Veilleux}, \&
  {Winter}}]{Koss:2010}
{Koss}, M., {Mushotzky}, R., {Veilleux}, S., \& {Winter}, L. 2010, \apjl, 716,
  L125

\bibitem[{{Koss} {et~al.}(2011){Koss}, {Mushotzky}, {Veilleux}, {Winter},
  {Baumgartner}, {Tueller}, {Gehrels}, \& {Valencic}}]{Koss:2011}
{Koss}, M., {Mushotzky}, R., {Veilleux}, S., {et~al.} 2011, \apj, 739, 57

\bibitem[{Koss {et~al.}(2011)Koss, Mushotzky, Treister, Veilleux, Vasudevan,
  Miller, Sanders, Schawinski, \& Trippe}]{Koss:2011:L42}
Koss, M., Mushotzky, R., Treister, E., {et~al.} 2011, \apjl, 735, L42,
  publisher: IOP Publishing

\bibitem[{{Koss} {et~al.}(2017){Koss}, {Trakhtenbrot}, {Ricci}, {Lamperti},
  {Oh}, {Berney}, {Schawinski}, {Balokovi{\'c}}, {Baronchelli}, {Crenshaw},
  {Fischer}, {Gehrels}, {Harrison}, {Hashimoto}, {Hogg}, {Ichikawa}, {Masetti},
  {Mushotzky}, {Sartori}, {Stern}, {Treister}, {Ueda}, {Veilleux}, \&
  {Winter}}]{Koss:2017}
{Koss}, M., {Trakhtenbrot}, B., {Ricci}, C., {et~al.} 2017, \apj, 850, 74

\bibitem[{{Koss} {et~al.}(2016){Koss}, {Assef}, {Balokovi{\'c}}, {Stern},
  {Gandhi}, {Lamperti}, {Alexander}, {Ballantyne}, {Bauer}, {Berney}, {Brandt},
  {Comastri}, {Gehrels}, {Harrison}, {Lansbury}, {Markwardt}, {Ricci},
  {Rivers}, {Schawinski}, {Trakhtenbrot}, {Treister}, \& {Urry}}]{Koss:2016}
{Koss}, M.~J., {Assef}, R., {Balokovi{\'c}}, M., {et~al.} 2016, \apj, 825, 85

\bibitem[{Koss {et~al.}(2018)Koss, Blecha, Bernhard, Hung, Lu, Trakthenbrot,
  Treister, Weigel, Sartori, Mushotzky, Schawinski, Ricci, Veilleux, \&
  Sanders}]{Koss:2018:214a}
Koss, M.~J., Blecha, L., Bernhard, P., {et~al.} 2018, \nat, 563, 214

\bibitem[{Koss {et~al.}(2021)Koss, Strittmatter, Lamperti, Shimizu,
  Trakhtenbrot, Saintonge, Treister, Cicone, Mushotzky, Oh, Ricci, Stern,
  Ananna, Bauer, Privon, Bär, De~Breuck, Harrison, Ichikawa, Powell, Rosario,
  Sanders, Schawinski, Shao, Megan~Urry, \& Veilleux}]{Koss:2021}
Koss, M.~J., Strittmatter, B., Lamperti, I., {et~al.} 2021, \apjs, 252, 29,
  publisher: IOP Publishing

\bibitem[{{Koss} {et~al.}(2022{\natexlab{a}}){Koss}, {Trakhtenbrot}, {Ricci},
  {Bauer}, {Treister}, {Mushotzky}, {Urry}, {Ananna}, {Balokovi{\'c}}, {den
  Brok}, {Cenko}, {Harrison}, {Ichikawa}, {Lamperti}, {Lein},
  {Mej{\'\i}a-Restrepo}, {Oh}, {Pacucci}, {Pfeifle}, {Powell}, {Privon},
  {Ricci}, {Salvato}, {Schawinski}, {Shimizu}, {Smith}, \&
  {Stern}}]{Koss:2021a}
{Koss}, M.~J., {Trakhtenbrot}, B., {Ricci}, C., {et~al.} 2022{\natexlab{a}},
  \apjs, 261, 1

\bibitem[{{Koss} {et~al.}(2022{\natexlab{b}}){Koss}, {Ricci}, {Trakhtenbrot},
  {Oh}, {den Brok}, {Mej{\'\i}a-Restrepo}, {Stern}, {Privon}, {Treister},
  {Powell}, {Mushotzky}, {Bauer}, {Ananna}, {Balokovi{\'c}}, {B{\"a}r},
  {Becker}, {Bessiere}, {Burtscher}, {Caglar}, {Congiu}, {Evans}, {Harrison},
  {Heida}, {Ichikawa}, {Kamraj}, {Lamperti}, {Pacucci}, {Ricci}, {Riffel},
  {Rojas}, {Schawinski}, {Temple}, {Urry}, {Veilleux}, \&
  {Williams}}]{Koss:2021b}
{Koss}, M.~J., {Ricci}, C., {Trakhtenbrot}, B., {et~al.} 2022{\natexlab{b}},
  \apjs, 261, 2

\bibitem[{{Koss} {et~al.}(2022{\natexlab{c}}){Koss}, {Trakhtenbrot}, {Ricci},
  {Oh}, {Bauer}, {Stern}, {Caglar}, {den Brok}, {Mushotzky}, {Ricci},
  {Mej{\'\i}a-Restrepo}, {Lamperti}, {Treister}, {B{\"a}r}, {Harrison},
  {Powell}, {Privon}, {Riffel}, {Rojas}, {Schawinski}, \& {Urry}}]{Koss:2022c}
{Koss}, M.~J., {Trakhtenbrot}, B., {Ricci}, C., {et~al.} 2022{\natexlab{c}},
  \apjs, 261, 6

\bibitem[{{Kravtsov} {et~al.}(2018){Kravtsov}, {Vikhlinin}, \&
  {Meshcheryakov}}]{Kravtsov:2018}
{Kravtsov}, A.~V., {Vikhlinin}, A.~A., \& {Meshcheryakov}, A.~V. 2018,
  Astronomy Letters, 44, 8

\bibitem[{{Krishnan} {et~al.}(2020){Krishnan}, {Almaini}, {Hatch}, {Wilkinson},
  {Maltby}, {Conselice}, {Kocevski}, {Suh}, \& {Wild}}]{Krishnan:2020}
{Krishnan}, C., {Almaini}, O., {Hatch}, N.~A., {et~al.} 2020, \mnras, 494, 1693

\bibitem[{{Krumpe} {et~al.}(2012){Krumpe}, {Miyaji}, {Coil}, \&
  {Aceves}}]{Krumpe:2012}
{Krumpe}, M., {Miyaji}, T., {Coil}, A.~L., \& {Aceves}, H. 2012, \apj, 746, 1

\bibitem[{{Krumpe} {et~al.}(2018){Krumpe}, {Miyaji}, {Coil}, \&
  {Aceves}}]{Krumpe:2017}
---. 2018, \mnras, 474, 1773

\bibitem[{{Krumpe} {et~al.}(2015){Krumpe}, {Miyaji}, {Husemann}, {Fanidakis},
  {Coil}, \& {Aceves}}]{Krumpe:2015}
{Krumpe}, M., {Miyaji}, T., {Husemann}, B., {et~al.} 2015, \apj, 815, 21

\bibitem[{{Lakhchaura} {et~al.}(2019){Lakhchaura}, {Truong}, \&
  {Werner}}]{Lakhchaura:2019}
{Lakhchaura}, K., {Truong}, N., \& {Werner}, N. 2019, \mnras, 488, L134

\bibitem[{{Landy} \& {Szalay}(1993)}]{Landy:1993}
{Landy}, S.~D., \& {Szalay}, A.~S. 1993, \apj, 412, 64

\bibitem[{{Lanzuisi} {et~al.}(2017){Lanzuisi}, {Delvecchio}, {Berta}, {Brusa},
  {Comastri}, {Gilli}, {Gruppioni}, {Marchesi}, {Perna}, {Pozzi}, {Salvato},
  {Symeonidis}, {Vignali}, {Vito}, {Volonteri}, \& {Zamorani}}]{Lanzuisi:2017}
{Lanzuisi}, G., {Delvecchio}, I., {Berta}, S., {et~al.} 2017, \aap, 602, A123

\bibitem[{{Laurent} {et~al.}(2017){Laurent}, {Eftekharzadeh}, {Le Goff},
  {Myers}, {Burtin}, {White}, {Ross}, {Tinker}, {Tojeiro}, {Bautista},
  {Brinkmann}, {Comparat}, {Dawson}, {du Mas des Bourboux}, {Kneib}, {McGreer},
  {Palanque- Delabrouille}, {Percival}, {Prada}, {Rossi}, {Schneider},
  {Weinberg}, {Y{\`e}che}, {Zarrouk}, \& {Zhao}}]{Laurent:2017}
{Laurent}, P., {Eftekharzadeh}, S., {Le Goff}, J.-M., {et~al.} 2017, Journal of
  Cosmology and Astro-Particle Physics, 2017, 017

\bibitem[{{Leauthaud} {et~al.}(2015){Leauthaud}, {J.~Benson}, {Civano},
  {L.~Coil}, {Bundy}, {Massey}, {Schramm}, {Schulze}, {Capak}, {Elvis},
  {Kulier}, \& {Rhodes}}]{Leauthaud:2015}
{Leauthaud}, A., {J.~Benson}, A., {Civano}, F., {et~al.} 2015, \mnras, 446,
  1874

\bibitem[{{Marasco} {et~al.}(2021){Marasco}, {Cresci}, {Posti}, {Fraternali},
  {Mannucci}, {Marconi}, {Belfiore}, \& {Fall}}]{Marasco:2021}
{Marasco}, A., {Cresci}, G., {Posti}, L., {et~al.} 2021, \mnras, 507, 4274

\bibitem[{{Marshall} {et~al.}(2018){Marshall}, {Shabala}, {Krause}, {Pimbblet},
  {Croton}, \& {Owers}}]{Marshall:2018}
{Marshall}, M.~A., {Shabala}, S.~S., {Krause}, M. G.~H., {et~al.} 2018, \mnras,
  474, 3615

\bibitem[{{Mazzucchelli} {et~al.}(2017){Mazzucchelli}, {Ba{\~n}ados},
  {Venemans}, {Decarli}, {Farina}, {Walter}, {Eilers}, {Rix}, {Simcoe},
  {Stern}, {Fan}, {Schlafly}, {De Rosa}, {Hennawi}, {Chambers}, {Greiner},
  {Burgett}, {Draper}, {Kaiser}, {Kudritzki}, {Magnier}, {Metcalfe}, {Waters},
  \& {Wainscoat}}]{Mazzucchelli:2017}
{Mazzucchelli}, C., {Ba{\~n}ados}, E., {Venemans}, B.~P., {et~al.} 2017, \apj,
  849, 91

\bibitem[{{Mej{\'\i}a-Restrepo} {et~al.}(2022){Mej{\'\i}a-Restrepo},
  {Trakhtenbrot}, {Koss}, {Oh}, {den Brok}, {Stern}, {Powell}, {Ricci},
  {Caglar}, {Ricci}, {Bauer}, {Treister}, {Harrison}, {Urry}, {Ananna},
  {Asmus}, {Assef}, {B{\"a}r}, {Bessiere}, {Burtscher}, {Ichikawa}, {Kakkad},
  {Kamraj}, {Mushotzky}, {Privon}, {Rojas}, {Sani}, {Schawinski}, \&
  {Veilleux}}]{Mejia-Restrepo:2022}
{Mej{\'\i}a-Restrepo}, J.~E., {Trakhtenbrot}, B., {Koss}, M.~J., {et~al.} 2022,
  \apjs, 261, 5

\bibitem[{{Mendez} {et~al.}(2016){Mendez}, {Coil}, {Aird}, {Skibba},
  {Diamond-Stanic}, {Moustakas}, {Blanton}, {Cool}, {Eisenstein}, {Wong}, \&
  {Zhu}}]{Mendez:2016}
{Mendez}, A.~J., {Coil}, A.~L., {Aird}, J., {et~al.} 2016, \apj, 821, 55

\bibitem[{{Meyer} {et~al.}(2022){Meyer}, {Decarli}, {Walter}, {Li}, {Wang},
  {Mazzucchelli}, {Ba{\~n}ados}, {Farina}, \& {Venemans}}]{Meyer:2022}
{Meyer}, R.~A., {Decarli}, R., {Walter}, F., {et~al.} 2022, \apj, 927, 141

\bibitem[{{Moster} {et~al.}(2013){Moster}, {Naab}, \& {White}}]{Moster:2013}
{Moster}, B.~P., {Naab}, T., \& {White}, S. D.~M. 2013, \mnras, 428, 3121

\bibitem[{Mushotzky {et~al.}(2014)Mushotzky, Shimizu, Meléndez, \&
  Koss}]{Mushotzky:2014:L34}
Mushotzky, R.~F., Shimizu, T.~T., Meléndez, M., \& Koss, M. 2014, \apjl, 781,
  L34

\bibitem[{{Oh} {et~al.}(2022){Oh}, {Koss}, {Ueda}, {Stern}, {Ricci},
  {Trakhtenbrot}, {Powell}, {den Brok}, {Lamperti}, {Mushotzky}, {Ricci},
  {B{\"a}r}, {Rojas}, {Ichikawa}, {Riffel}, {Treister}, {Harrison}, {Urry},
  {Bauer}, \& {Schawinski}}]{Oh:2021}
{Oh}, K., {Koss}, M.~J., {Ueda}, Y., {et~al.} 2022, \apjs, 261, 4

\bibitem[{{Planck Collaboration} {et~al.}(2016){Planck Collaboration}, {Ade},
  {Aghanim}, {Arnaud}, {Ashdown}, {Aumont}, {Baccigalupi}, {Banday},
  {Barreiro}, {Bartlett}, \& et~al.}]{Planck:2015}
{Planck Collaboration}, {Ade}, P.~A.~R., {Aghanim}, N., {et~al.} 2016, \aap,
  594, A13

\bibitem[{{Powell} {et~al.}(2020){Powell}, {Urry}, {Cappelluti}, {Johnson},
  {LaMassa}, {Ananna}, \& {Kollmann}}]{Powell:2020}
{Powell}, M.~C., {Urry}, C.~M., {Cappelluti}, N., {et~al.} 2020, \apj, 891, 41

\bibitem[{{Powell} {et~al.}(2017){Powell}, {Urry}, {Cardamone}, {Simmons},
  {Schawinski}, {Young}, \& {Kawakatsu}}]{Powell:2017}
{Powell}, M.~C., {Urry}, C.~M., {Cardamone}, C.~N., {et~al.} 2017, \apj, 835,
  22

\bibitem[{{Powell} {et~al.}(2018){Powell}, {Cappelluti}, {Urry}, {Koss},
  {Finoguenov}, {Ricci}, {Trakhtenbrot}, {Allevato}, {Ajello}, {Oh},
  {Schawinski}, \& {Secrest}}]{Powell:2018}
{Powell}, M.~C., {Cappelluti}, N., {Urry}, C.~M., {et~al.} 2018, \apj, 858, 110

\bibitem[{{Reines} \& {Volonteri}(2015)}]{Reines:2015}
{Reines}, A.~E., \& {Volonteri}, M. 2015, \apj, 813, 82

\bibitem[{{Ricarte} {et~al.}(2020){Ricarte}, {Tremmel}, {Natarajan}, \&
  {Quinn}}]{Ricarte:2020}
{Ricarte}, A., {Tremmel}, M., {Natarajan}, P., \& {Quinn}, T. 2020, \apjl, 895,
  L8

\bibitem[{{Ricci} {et~al.}(2015){Ricci}, {Ueda}, {Koss}, {Trakhtenbrot},
  {Bauer}, \& {Gandhi}}]{Ricci:2015}
{Ricci}, C., {Ueda}, Y., {Koss}, M.~J., {et~al.} 2015, \apjl, 815, L13

\bibitem[{{Ricci} {et~al.}(2017){Ricci}, {Trakhtenbrot}, {Koss}, {Ueda}, {Del
  Vecchio}, {Treister}, {Schawinski}, {Paltani}, {Oh}, {Lamperti}, {Berney},
  {Gandhi}, {Ichikawa}, {Bauer}, {Ho}, {Asmus}, {Beckmann}, {Soldi},
  {Balokovi{\'c}}, {Gehrels}, \& {Markwardt}}]{Ricci:2017B}
{Ricci}, C., {Trakhtenbrot}, B., {Koss}, M.~J., {et~al.} 2017, \apjs, 233, 17

\bibitem[{{Robinson} {et~al.}(2021){Robinson}, {Bentz}, {Courtois}, {Johnson},
  {Crenshaw}, {Meena}, {Polack}, {Silverstein}, \& {Chen}}]{Robinson:2021}
{Robinson}, J.~H., {Bentz}, M.~C., {Courtois}, H.~M., {et~al.} 2021, \apj, 912,
  160

\bibitem[{{Sabra} {et~al.}(2015){Sabra}, {Saliba}, {Abi Akl}, \&
  {Chahine}}]{Sabra:2015}
{Sabra}, B.~M., {Saliba}, C., {Abi Akl}, M., \& {Chahine}, G. 2015, \apj, 803,
  5

\bibitem[{{Saha} \& {Naab}(2013)}]{Saha:2013}
{Saha}, K., \& {Naab}, T. 2013, \mnras, 434, 1287

\bibitem[{{Shankar} {et~al.}(2016){Shankar}, {Bernardi}, {Sheth}, {Ferrarese},
  {Graham}, {Savorgnan}, {Allevato}, {Marconi}, {L{\"a}sker}, \&
  {Lapi}}]{Shankar:2016}
{Shankar}, F., {Bernardi}, M., {Sheth}, R.~K., {et~al.} 2016, \mnras, 460, 3119

\bibitem[{{Shankar} {et~al.}(2020){Shankar}, {Allevato}, {Bernardi}, {Marsden},
  {Lapi}, {Menci}, {Grylls}, {Krumpe}, {Zanisi}, {Ricci}, {La Franca}, {Baldi},
  {Moreno}, \& {Sheth}}]{Shankar:2020}
{Shankar}, F., {Allevato}, V., {Bernardi}, M., {et~al.} 2020, Nature Astronomy,
  4, 282

\bibitem[{{Shen} {et~al.}(2009){Shen}, {Strauss}, {Ross}, {Hall}, {Lin},
  {Richards}, {Schneider}, {Weinberg}, {Connolly}, {Fan}, {Hennawi}, {Shankar},
  {Vanden Berk}, {Bahcall}, \& {Brunner}}]{Shen:2009}
{Shen}, Y., {Strauss}, M.~A., {Ross}, N.~P., {et~al.} 2009, \apj, 697, 1656

\bibitem[{{Shimizu} {et~al.}(2017){Shimizu}, {Mushotzky}, {Mel{\'e}ndez},
  {Koss}, {Barger}, \& {Cowie}}]{Shimizu:2017}
{Shimizu}, T.~T., {Mushotzky}, R.~F., {Mel{\'e}ndez}, M., {et~al.} 2017,
  \mnras, 466, 3161

\bibitem[{{Simola} {et~al.}(2019){Simola}, {Cisewski-Kehe}, {Gutmann}, \&
  {Corander}}]{Simola:2019}
{Simola}, U., {Cisewski-Kehe}, J., {Gutmann}, M.~U., \& {Corander}, J. 2019,
  arXiv e-prints, arXiv:1907.01505

\bibitem[{{Smith} {et~al.}(2021){Smith}, {Bureau}, {Davis}, {Cappellari},
  {Liu}, {Onishi}, {Iguchi}, {North}, \& {Sarzi}}]{Smith:2021}
{Smith}, M.~D., {Bureau}, M., {Davis}, T.~A., {et~al.} 2021, \mnras, 500, 1933

\bibitem[{{Starikova} {et~al.}(2011){Starikova}, {Cool}, {Eisenstein},
  {Forman}, {Jones}, {Hickox}, {Kenter}, {Kochanek}, {Kravtsov}, {Murray}, \&
  {Vikhlinin}}]{Starikova:2011}
{Starikova}, S., {Cool}, R., {Eisenstein}, D., {et~al.} 2011, \apj, 741, 15

\bibitem[{{Suh} {et~al.}(2020){Suh}, {Civano}, {Trakhtenbrot}, {Shankar},
  {Hasinger}, {Sanders}, \& {Allevato}}]{Suh:2020}
{Suh}, H., {Civano}, F., {Trakhtenbrot}, B., {et~al.} 2020, \apj, 889, 32

\bibitem[{{Suh} {et~al.}(2019){Suh}, {Civano}, {Hasinger}, {Lusso}, {Marchesi},
  {Schulze}, {Onodera}, {Rosario}, \& {Sanders}}]{Suh:2019}
{Suh}, H., {Civano}, F., {Hasinger}, G., {et~al.} 2019, \apj, 872, 168

\bibitem[{{Timlin} {et~al.}(2018){Timlin}, {Ross}, {Richards}, {Myers},
  {Pellegrino}, {Bauer}, {Lacy}, {Schneider}, {Wollack}, \&
  {Zakamska}}]{Timlin:2018}
{Timlin}, J.~D., {Ross}, N.~P., {Richards}, G.~T., {et~al.} 2018, \apj, 859, 20

\bibitem[{{Tinker} {et~al.}(2010){Tinker}, {Robertson}, {Kravtsov}, {Klypin},
  {Warren}, {Yepes}, \& {Gottl{\"o}ber}}]{Tinker:2010}
{Tinker}, J.~L., {Robertson}, B.~E., {Kravtsov}, A.~V., {et~al.} 2010, \apj,
  724, 878

\bibitem[{{van den Bosch} {et~al.}(2016){van den Bosch}, {Jiang}, {Campbell},
  \& {Behroozi}}]{vandenBosch:2016}
{van den Bosch}, F.~C., {Jiang}, F., {Campbell}, D., \& {Behroozi}, P. 2016,
  \mnras, 455, 158

\bibitem[{{Vasudevan} \& {Fabian}(2007)}]{Vasudevan:2007}
{Vasudevan}, R.~V., \& {Fabian}, A.~C. 2007, \mnras, 381, 1235

\bibitem[{{Viitanen} {et~al.}(2021){Viitanen}, {Allevato}, {Finoguenov},
  {Shankar}, \& {Marsden}}]{Viitanen:2021}
{Viitanen}, A., {Allevato}, V., {Finoguenov}, A., {Shankar}, F., \& {Marsden},
  C. 2021, \mnras, 507, 6148

\bibitem[{{Weinberger} {et~al.}(2018){Weinberger}, {Springel}, {Pakmor},
  {Nelson}, {Genel}, {Pillepich}, {Vogelsberger}, {Marinacci}, {Naiman},
  {Torrey}, \& {Hernquist}}]{Weinberger:2018}
{Weinberger}, R., {Springel}, V., {Pakmor}, R., {et~al.} 2018, \mnras, 479,
  4056

\bibitem[{{White} {et~al.}(2012){White}, {Myers}, {Ross}, {Schlegel},
  {Hennawi}, {Shen}, {McGreer}, {Strauss}, {Bolton}, {Bovy}, {Fan},
  {Miralda-Escude}, {Palanque-Delabrouille}, {Paris}, {Petitjean}, {Schneider},
  {Viel}, {Weinberg}, {Yeche}, {Zehavi}, {Pan}, {Snedden}, {Bizyaev},
  {Brewington}, {Brinkmann}, {Malanushenko}, {Malanushenko}, {Oravetz},
  {Simmons}, {Sheldon}, \& {Weaver}}]{White:2012}
{White}, M., {Myers}, A.~D., {Ross}, N.~P., {et~al.} 2012, \mnras, 424, 933

\bibitem[{{Wise} {et~al.}(2019){Wise}, {Regan}, {O'Shea}, {Norman}, {Downes},
  \& {Xu}}]{Wise:2019}
{Wise}, J.~H., {Regan}, J.~A., {O'Shea}, B.~W., {et~al.} 2019, \nat, 566, 85

\bibitem[{{Yang} {et~al.}(2018){Yang}, {Brandt}, {Darvish}, {Chen}, {Vito},
  {Alexander}, {Bauer}, \& {Trump}}]{Yang:2018}
{Yang}, G., {Brandt}, W.~N., {Darvish}, B., {et~al.} 2018, \mnras, 480, 1022

\end{thebibliography}
\bibliographystyle{aasjournal}

\appendix
\section{Impact of Different \Mbh\ Measurement Methods}
\label{sec:a1}
There are various methods for measuring SMBH mass, which largely depend on whether broad lines are present in the optical spectra. The hard-X-ray sensitivity of {\it Swift}/BAT detects both Type 1 (unobscured) and Type 2 (obscured) AGN, and so multiple methods are used in our sample.

Type 2 methods rely on the local $M-\sigma_{*}$ relation and are typically understood to have larger uncertainties than broad line estimates utilized for Type 1 AGN (\citealt{Mejia-Restrepo:2022,Koss:2022c}). It has also been proposed that they may be systematically biased due to selection effects \citep{Shankar:2016}.

\begin{figure}[h]
    \centering
    \includegraphics[width=.49\textwidth]{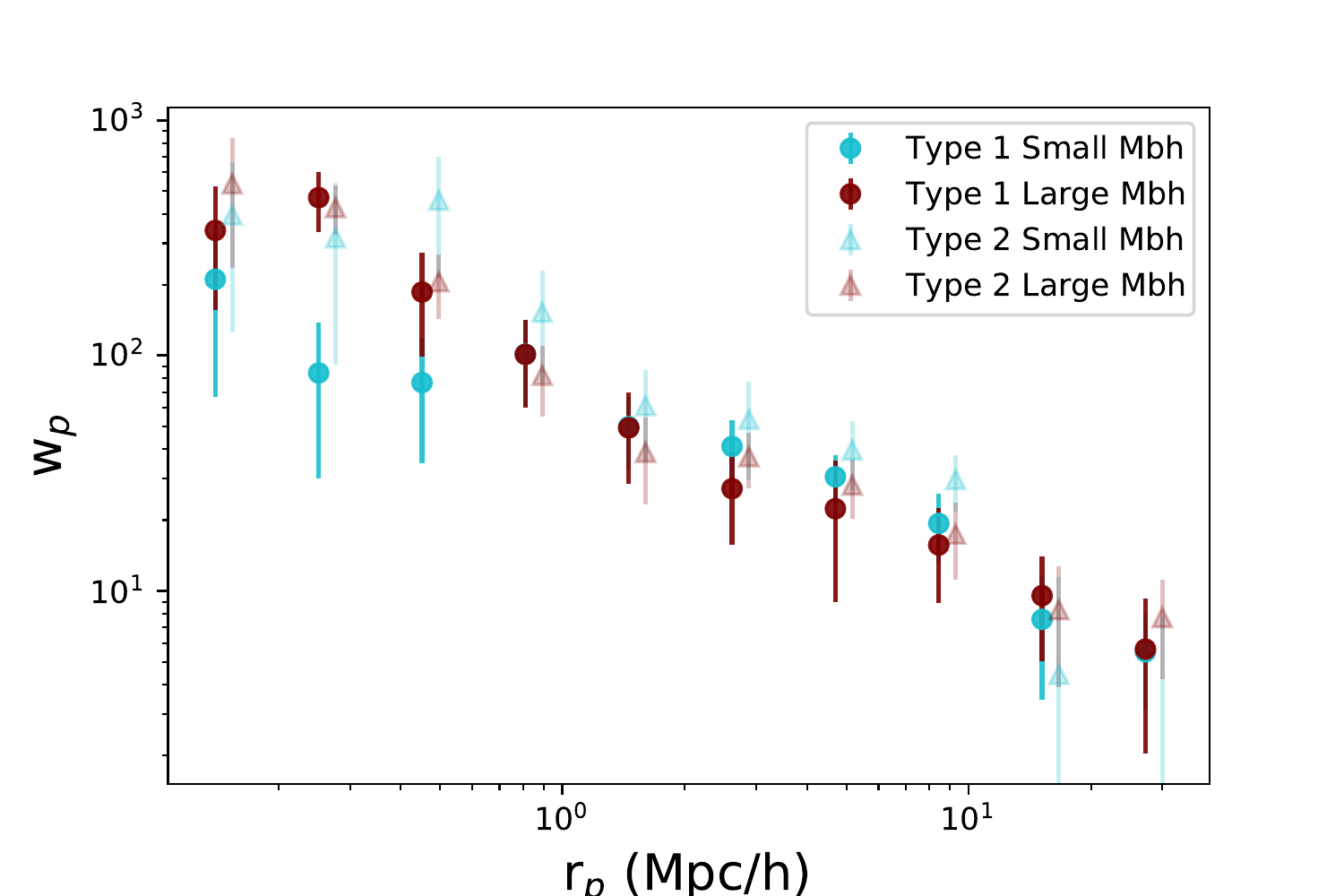}
    \includegraphics[width=.49\textwidth]{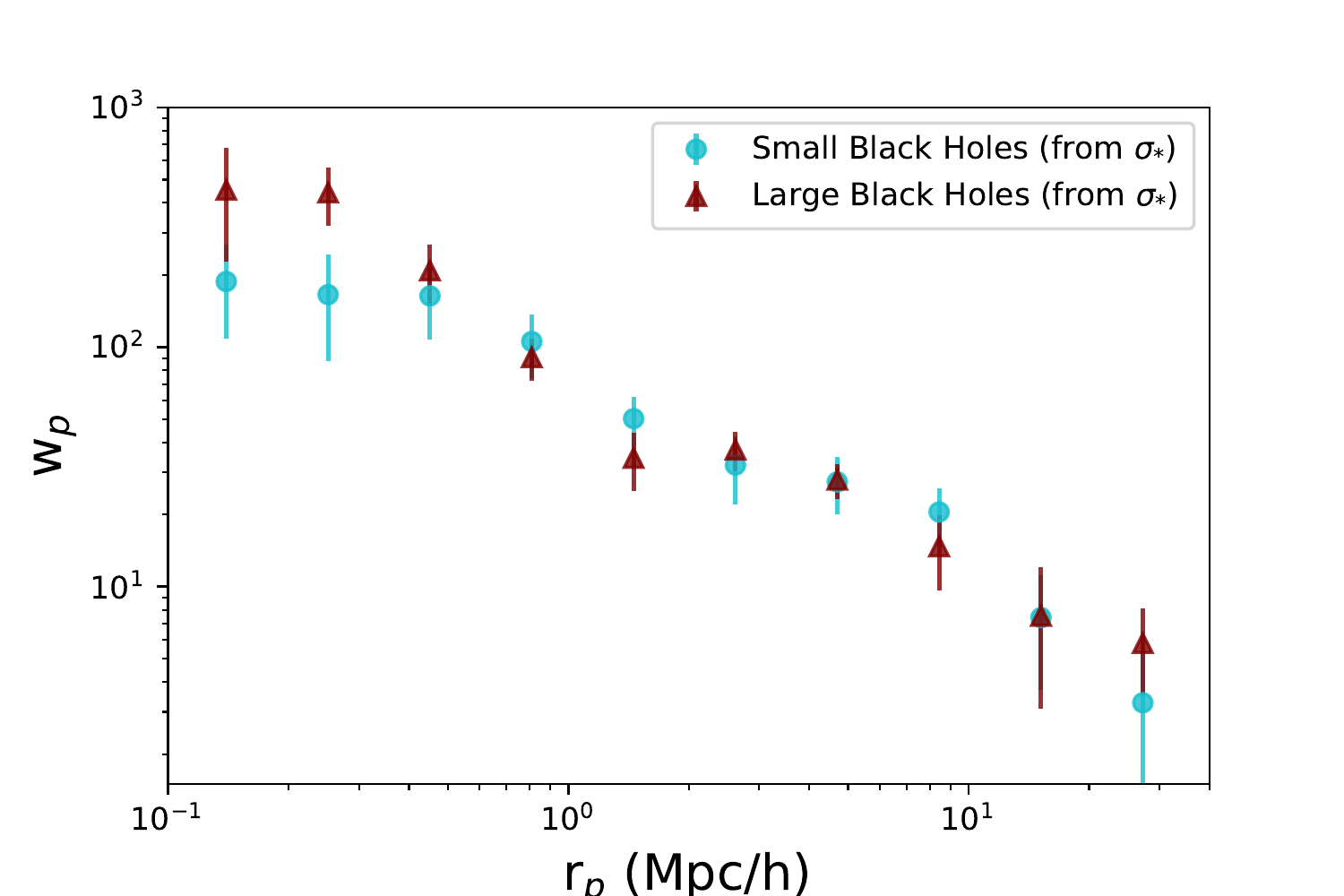}
    \caption{{\it Left}: Projected correlation functions of each \Mbh\ bin (small \Mbh, cyan; large \Mbh, maroon; same bins as Fig. \ref{fig:mbh_vs_z}) disagreggated by the two main mass estimation methods. Type 1 AGN, whose \Mbh\ are estimated by broad-line measurements, are shown by the filled circles. Type 2 AGN, whose mass estimates rely on the \Mbh$-\sigma_{*}$ relation, are shown by the lighter triangles. {\it Right}: Projected correlation functions of each \Mbh\ bin using black hole masses calculated solely with velocity dispersion measurements.}
    \label{fig:mbht1t2}
\end{figure}

\citealt{Powell:2018} found that obscured AGN clustered more strongly than unobscured AGN when controlling for AGN luminosity, redshift, and stellar mass. To test whether the black hole mass differences measured in this work are due to obscuration rather than black hole mass, we computed the clustering of large and small \Mbh\ for Type 1 and Type 2 AGN separately. Figure \ref{fig:mbht1t2} (left) shows the correlation functions of Type 1 (including types 1.2, 1.5, and 1.8) and Type 2 AGN for the two \Mbh\ bins. For the Type 1 AGN, there are still significant differences between the two mass bins on 1-halo term scales. For Type 2 AGN, the measurements are noisier and show no significant differences. 

To investigate further, we calculated the correlation functions of each \Mbh bin using black hole mass measurements obtained from stellar velocity dispersion (i.e., the Type 2 method; Caglar et al., {\it in prep.}), for all AGN types (Fig. \ref{fig:mbht1t2}, right). AGN with larger \Mbh\ were still found to cluster more strongly than AGN with smaller \Mbh.

Because each black hole estimation method showed a clustering difference on 1-halo scales between \Mbh\ bins, we conclude that obscuration is not predominantly driving the AGN clustering mass dependence.

\section{Sensitivity of ERDF}
The Eddington Ratio Distribution Function used in this work and constrained by \citealt{Ananna:2022} was parameterized in the following manner:

\begin{equation}
    \frac{dN}{d \log \lambda_{\rm Edd}} \propto \left[ \left(\frac{\lambda_{\rm Edd}}{\lambda_{Edd}^{*}}\right)^{\delta_{1}} + \left(\frac{\lambda_{\rm Edd}}{\lambda_{\rm Edd}^{*}}\right)^{\delta_{2}} \right]^{-1}
\end{equation}
    
\noindent with three parameters (characteristic Eddington ratio, $\lambda_{\rm Edd}^{*}$; low-Eddington slope, $\delta_{1}$; and high-Eddington slope, $\delta_{2}$, where $\delta_{2}>\delta_1$). Our analysis marginalized over these parameters using the constraints from \citealt{Ananna:2022}. 

\begin{figure}
    \centering
    \includegraphics[width=.47\textwidth]{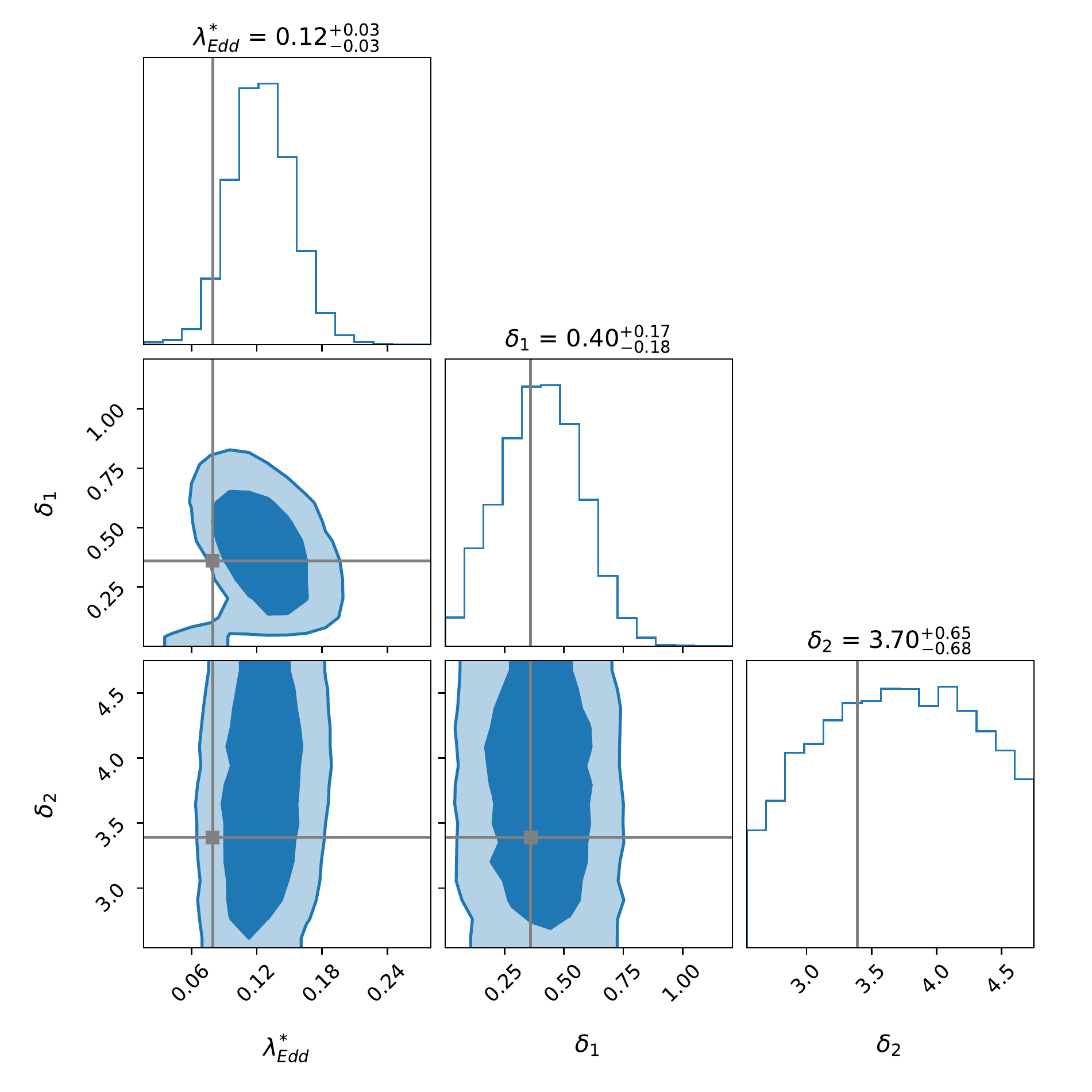}
    \includegraphics[width=.47\textwidth]{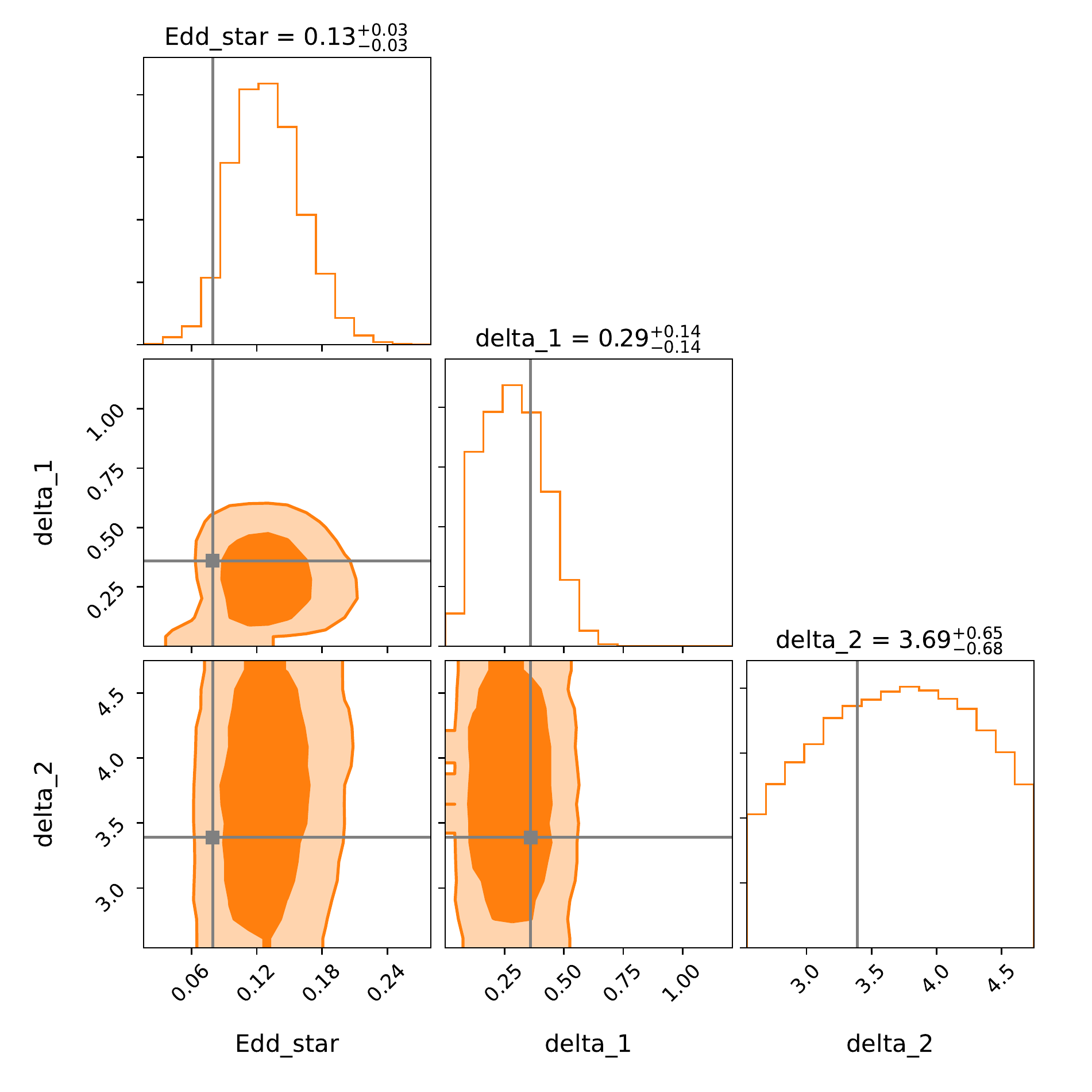}
    \caption{Corner plots of the covariances of ERDF parameters}
    \label{fig:erdf_params}
\end{figure}

To investigate the sensitivity of these ERDF parameters, we repeated our analysis fixing $M_{\rm{BH}}-M_{*}$ parameters and varying the ERDF parameters with larger, top-hat priors. We tested whether consistent ERDF parameters are found. Figure \ref{fig:erdf_params} shows the corner plots of the ABC analysis varying ERDF parameters for fixed normalization, scatter, and slope of $M_{\rm{BH}}-M_{*}$.

While $\delta_2$ was unconstrained by the analysis (and therefore insensitive to the results), $\delta_1$ and $\lambda_{Edd}$ were found to be consistent with within the uncertainties of the fiducial values. 

\section{Effect of SMHR Parameters}
\label{sec:a3}
For our analysis, we assume the $M_{*}-M_{peak}$ relation (SHMR) from \citealt{Behroozi:2010}. 
However, there are uncertainties in the parameters of this relation. Since the clustering and space density measurements are sensitive to the \Mbh$-M_{halo}$ connection, changing the SHMR parameters impacts the resulting \Mbh$-M_{*}$ parameters that we constrain.

We reran our analysis assuming a smaller scatter (0.15 dex, for fixed halo mass) in this relation (as is reported in \citealt{Moster:2013}), which resulted in slightly higher values for the scatter on the \Mbh$-M_{*}$ relation (i.e., $0.35\pm0.17$ and $0.45\pm0.18$ for Model 1 and 2, respectively), as well as slightly lower normalizations ($7.72\pm0.30$, $7.54\pm0.36$) and slopes ($0.62\pm0.24$, $0.51\pm0.23$). 

There also have been SHMR measurements with different slopes than those reported in \citealt{Behroozi:2010}, especially on the high-mass end \citep[e.g.,][]{Moster:2013,Kravtsov:2018}. We repeated our analysis using the \citealt{Moster:2013} mean SMHR relation (which features a steeper high-mass slope). This resulted in the following best-fit \Mbh$-M_{*}$ parameters: a normalization of $7.69\pm0.3$ ($7.49\pm0.35$), a slope of $0.66\pm0.24$ ($0.47\pm$0.23), and a scatter of $0.35\pm0.17$ ($0.43\pm$0.17) for Model 1 (2). Model 2 was still preferred in each case.

\end{document}